# Enabling Heterogeneous Catalysis to Achieve Carbon Neutrality: Directional Catalytic Conversion of CO$_2$ into Carboxylic Acids

*Xiaofei Zhang, xxx, Huabin Zhang\**

[*] Dr. X. Zhang, Prof. H. Zhang

KAUST Catalysis Center (KCC), King Abdullah University of Science and Technology (KAUST), Thuwal 23955-6900, Saudi Arabia

Email: Huabin.zhang@kaust.edu.sa

## Abstract

The increase in anthropogenic carbon dioxide (CO$_2$) emissions has exacerbated the deterioration of the global environment, which should be controlled to achieve carbon neutrality. Central to the core goal of achieving carbon neutrality is the utilization of CO$_2$ under economic and sustainable conditions. Recently, the strong need for carbon neutrality has led to a proliferation of studies on the direct conversion of CO$_2$ into carboxylic acids, which could effectively alleviate CO$_2$ emissions and create high-value chemicals. The purpose of this review is to present the application prospects of carboxylic acids and the basic principles of CO$_2$ conversion into carboxylic acids through photo-, electric-, and thermal catalysis. Special attention is focused on the regulation strategy of the activity of abundant catalysts at the molecular level, inspiring the preparation of high-performance catalysts. In addition, theoretical calculation, advanced technologies, and numerous typical examples are introduced to elaborate on the corresponding process and influencing factors of catalytic activity. Finally, challenges and prospects are provided for the future development of this field. It is hoped that this review contributes to a deeper understanding of the conversion of $CO_2$ into carboxylic acids and inspires more innovative breakthroughs.

# 1. Introduction

## 1.1 General Background

With the intensification of the global energy crisis and environmental deterioration caused by the warming climate, aiming for carbon-neutral recycling is in imminent demand regardless of environmental protection or economic development(*1, 2*). The effective conversion of carbon dioxide ($CO_2$) into fine chemicals has undoubtedly emerged as a powerful strategy to achieve carbon-neutral recycling due to its low cost and potentially low energy consumption(*3-5*). Extensive cost-effective conversion paths have been explored to realize such conversion practicably(*5-8*). To date, various methods and strategies, including biological transformation, photocatalytic reduction, electrocatalytic reduction, organic transformation, dry reforming, and others, have been explored to convert $CO_2$ into valuable chemicals based on continuing in-depth studies over the past two decades(*9-11*).

In general, $CO_2$ can be converted into numerous chemicals, such as CO, carboxylic acids, $CH_3OH$, $CH_4$, olefin, and so on, under the drivers of light, electricity, and heat(*12, 13*). A large volume of $CO_2$ emissions in the atmosphere is regarded as exhaust gas and the root cause of environmental deterioration, whereas these products could become valuable raw chemical materials with wide application value through catalytic conversion. A well-known application of CO is the synthesis of aldehydes through hydroformylation reaction, and high toxicity also poses challenges for its use and storage(*14*). Another product, $CH_3OH$, can be widely used as a fuel and is a vital chemical compound for producing long-chain hydrocarbons(*15*). Similarly, methane is an important storage fuel with the advantage of low-cost storage(*16*). Produced olefins, such as ethylene and propylene, are key chemicals in fabricating plastics, medicines, and paints(*17*). Although $CH_3OH$, $CH_4$, and olefin have obvious economic value, their production processes undergo multiple electron transfers, leading to poor conversion efficiency and unsatisfactory product selectivity(*18*). In contrast to these products, carboxylic acids occupy a unique position due to the great breakthroughs made in recent years and their commercial value and apparent advantages, such as low toxicity, high density, and high value per kWh of electrical energy input, over other products.

Despite the great progress achieved in the conversion and utilization of $CO_2$ in the past few decades, the extreme stability of $CO_2$ creates considerable challenges in its activation to participate in synthesizing carboxylic acids via specific intermediates and control the selectivity of carboxylic acids due to competitive reactions. This daunting challenge has prompted the development of outstanding heterogeneous catalysts to efficiently convert $CO_2$ into carboxylic acids under mild conditions.

Currently, the reported emerging nano-materials, including clusters, metal oxide, porous materials, organic-inorganic hybrid materials and alloys, have proved efficient heterogeneous catalysts presenting superior catalytic activity to homogeneous catalysts in producing carboxylic acids(*19, 20*). More effort should be made to discern the decisive factors in producing carboxylic acids through advanced characterization techniques and theoretical calculations to develop ideal catalysts based on an accurate understanding of the mechanical process to realize large-scale, efficient production of carboxylic acids with $CO_2$.

## 1.2 Introduction of Carboxylic Acid Compounds

Carboxylic acids encompassing formic, acetic, benzoic, acetylenic, amino, lactic, and other acids are essential components in fine chemicals and are regarded as core compounds in the natural carbon cycle(*21*). These chemicals have great application value in synthesizing chemical compounds and energy conversion. In all these carboxylic acids, formic acid has the smallest molecular formula and can be obtained in high yield with clean water or hydrogen as a reactant to react with $CO_2$ over numerous heterogeneous catalysts(*22*). In addition, it is also regarded as a safe and convenient hydrogen storage carrier due to its high volumetric capacity of 53 g $H_2$/L, which can be converted into clean electricity in direct formic acid fuel cells with appealing advantages, such as easy transportation and storage and high theoretical open-circuit voltage(*23*). Furthermore, it is an important chemical intermediate, which can be widely used in the leather, pesticide, antibacterial agent, medicine, livestock feed, dyestuff, and rubber industries(*24*). Moreover, other multi-carbon carboxylic acids have equivalent application prospects and economic value to formic acid. For example, they have been widely used as preservatives in pharmaceutical, cosmetic, agriculture, and polymer industries and are recognized as valuable platform chemicals for a growing nonfossil industry(*25*).

The industrial production of carboxylic acids has undergone a cumbersome and ineffective process that generates more energy consumption and threatens the current environment. For instance, one adopted route to synthesize formic acid undergoes the direct carboxylation of methanol and subsequent hydrolysis of methyl formate, which has evident shortcomings, such as high cost and low production efficiency(*26*). Another developed synthesis route using lignocellulosic biomass to produce formic acid still suffers from complicated steps, such as acid hydrolysis, wet oxidation, and catalytic oxidation, leading to ultra-low biomass utilization and poor selectivity(*27*). Recently, the direct conversion of $CO_2$ into formic acid has become widely appealing due to the drive for carbon neutrality and the economics of the synthetic route. The high yield and selectivity of formic acid are expected under a low energy consumption, profiting from a one-step synthesis path and well-developed heterogeneous $CO_2$ activation process.

Likewise, the synthesis of other carboxylic acids, such as benzoic, acetylenic, amino, and lactic acids, faces vigorous challenges in activating inert $CO_2$ and C-H bonds of organic substrates, leading to excessive energy consumption and possible impure products from competitive reactions(*28*). This outcome is evident in the synthesis of benzoic acid by the partial oxidation of toluene using a cobalt-manganese catalyst under high temperature. Benzaldehyde forms as a side product, and the conversion rates of toluene are far from satisfactory. Another case is the preparation of acetylenic acid, where metal-organic reagents that are sensitive to water and air are often needed and operated under anhydrous and oxygen-free conditions, increasing the production cost and creating challenges for industrialized operations(*29*). In the case of producing amino and lactic acids, the original organic reactants containing multiple functional groups have the possibility of side reactions. The transformation of functional groups that may produce undesired products should be avoided.

The central goal is to prepare various carboxylic acids with $CO_2$ with a high conversion rate and selectivity under mild conditions to overcome the above challenges and minimize carbon emissions. Thus far, significant progress has been reached in producing numerous kinds of carboxylic acids through photocatalytic reduction, electrocatalytic reduction, and thermally driven organic transformation (Figure 1)(*30-32*). Despite the great achievements, the unsatisfactory utilization rate of catalytic sites, competitive side reactions, and unclear reaction mechanism pathways should still be addressed by developing efficient heterogeneous catalysts and revealing the inner relationship between the structure and activity.

**1.3 Scope of this Review**

The increasing literatures recognize the urgency of reducing $CO_2$ emissions to achieve carbon neutrality. Most published reviews are focused on the specific materials and characterization techniques in catalyzing $CO_2$ conversion or are limited to preparing formic acid from $CO_2$*(33-36)*. However, few reviews have systematically summarized the progress and challenges in producing carboxylic acids using $CO_2$. A systematic understanding of how the catalyst structure contributes to the yield or selectivity of products is still lacking, and it is not clear which factors primarily contribute to the conversion efficiency and product selectivity. Therefore, an opportune overview of construction strategies for advanced catalysts and an elaboration of the mechanical process in converting $CO_2$ into carboxylic acids through the drivers of light, electricity, and heat are highly desirable and may provide important insight into advancing the development of practical strategies in mitigating carbon emissions and producing more high-value chemicals, such as carboxylic acids.

This review starts with a general introduction of the basic principles of catalyzing $CO_2$ into various carboxylic acids through photocatalysis, electrocatalysis, and thermal catalysis. Then, we highlight the adopted strategies, including composite catalysts, heteroatom doping, morphology control, and surface functionalization in rationally engineering heterogeneous catalysts to boost the conversion of $CO_2$ into carboxylic acids. Furthermore, recent experimental and theoretical progress has revealed the reaction mechanism and presented rich examples to reveal the achievements in converting $CO_2$ to carboxylic acids. Finally, the key challenges and prospects are presented for guiding the rationale design of heterogeneous catalysts and the improvement in the conversion efficiency of $CO_2$ to carboxylic acids. We aim to illustrate the important relationship between the structure of the designed catalysts and reaction activity by discussing vital influencing parameters in the reaction process and providing fundamental principles and strategies for further developing superior heterogeneous catalysts.

## 2. Fundamentals of Catalyzing $CO_2$ into Carboxylic Acid

### 2.1 Photocatalytic $CO_2$ Conversion Process

Photocatalytic reduction of $CO_2$ into highly valuable chemicals has been attracting considerable attention due to the economy of solar energy, which is expected to replace fossil fuels, and the utilization of $CO_2$ with minimal carbon emissions (Figure 2A). In this process, well-designed photocatalysts with specific bandgap sizes and positions of the conduction band (CB) and valence band (VB) occupy a core position in enhancing solar energy utilization due to their ability to harvest light and drive the following redox reactions(*37, 38*). Typically, the photocatalytic $CO_2$ reduction to carboxylic acids includes the following three steps: (1) light-harvesting, (2) generation and separation of a photogenerated charge carrier, and (3) surface redox reactions (Figure 2B). The corresponding photocatalytic process involves harvesting the incident photons by generating photoexcited electrons ($e^-$) and holes ($h^+$) over the traditional semiconductors. Light-harvesting determines the utilization rate of solar energy and facilitates the following $CO_2$ photoreduction step. Considering that visible light and near-infrared (IR) light account for more than 95% of natural light, many attempts have been made to enhance light-harvesting photocatalysts(*39*). For example, the visible-light absorption of graphitic carbon nitride can be increased due to the produced nitrogen vacancies and modified electron structure through polymerization of dicyandiamide using tartaric acid(*40*). Furthermore, another example demonstrates that the enhanced photocatalysis performance can be realized in Ti-doped $SnS_2$ by forming an intermediate bond to extend the absorption of IR light(*41*).

In the second step, the formed electron and hole (e-h) pairs migrate to the surface of the designed catalysts. The photogenerated charge separation process is accompanied by recombination, and only a few photogenerated electrons participate in the following $CO_2$-involved reaction. Numerous strategies have been developed to suppress the recombination of charge carriers to overcome the inefficient utilization of photogenerated electrons. For instance, the photocatalytic activity of $Bi_2WO_6$ increased greatly by reducing the bulk $Bi_2WO_6$ into $Bi_2WO_6$ nanosheets (NSs) due to the minimized charge transport pathway, leading to the rapid migration of photoexcited electrons and holes to the surface of photocatalysts(*42*). In addition, combining the semiconductor photocatalyst with carbon-based materials to form composite catalysts efficiently boosts the carrier separation efficiency due to the improved conductivity with the aid of support. Composite catalysts, such as CdS@biomass porous carbons, can be used as electron reservoirs to boost the separation of charge carriers and the transportation of electrons owing to good conductivity, leading to excellent activity in $CO_2$ photoreduction under visible irradiation(*43*).

After the charge migrates to the catalyst surface, it reacts with the $CO_2$ adsorbed around the active site. Engineering active sites can stimulate the occurrence of surface redox reaction and effectively regulate the activation energy barrier of the intermediates to produce carboxylic acids, leading to the desired products with high activity and selectivity(*44, 45*). For instance, the selectivity of formic acid can be tuned by installing different metal active centers with variable electron structures into covalent organic frameworks (COF). The metal sites, as a poor π-donor, prefer to enhance the C-O bonding force of $CO_2$, producing formic acid, whereas metal sites with an electron-rich coordination environment tend to weaken and break the C-O bond to form CO(*46*).

In the past few years, great progress has been made in boosting the conversion efficiency of $CO_2$ into carboxylic acids through photocatalysis by optimizing the structure and composition of photocatalysts, such as adjusting the band structure and allowing the electron pumps to accelerate the surface charge separation and transfer(*47*). Numerous highly valuable carboxylic acids can be obtained through the photocatalytic conversion path with $CO_2$ (Figure 2C). Nevertheless, the process of producing carboxylic acids is often accompanied by producing a diverse range of products via proton-coupled multi-electron transfer processes with thermodynamic potentials between about -0.20 and -0.70 V. For instance, formic acid is produced through two electron transfer processes with a potential of -0.61 V, which faces competitive reactions, such as the hydrogen evolution reaction with a potential of -0.42 V, reduction of $CO_2$ into $CH_4$ with a potential of -0.24 V, and reduction of $CO_2$ into CO with a potential of -0.52 V(*48*).

The unfavorable energy barrier for producing formic acid should be reduced by introducing efficient catalysts to enhance the reaction kinetics. Furthermore, some effective strategies, such as enhancing the hydrophobicity of the catalyst surface to inhibit the hydrogen evolution reaction and controlling the composition of the catalyst to suppress the occurrence of multiple electron transfer processes, should be considered to improve the selectivity of formic acid.

**2.2 Electrocatalytic $CO_2$ Conversion Process**

Electrocatalytic $CO_2$ reduction to valuable fuels represents an appealing way to make full use of $CO_2$ as a $C_1$ resource, driven by electricity generated through renewable energy sources, such as wind and hydropower (Figure 2D)(*49*). As a result of the significant decline in the cost of electricity, electrocatalytic $CO_2$ reduction exhibits great potential for future practical and large-scale industrial applications. The appealing advantages in this process include the highly selective preparation of chemicals, satisfactory electric-power utilization efficiency, and environmentally friendly reaction medium. Therefore, researchers are enthusiastic about exploring advanced devices, efficient catalysts, and green additives to realize high-efficiency electrocatalytic $CO_2$.

The design of heterogeneous catalysts for electrocatalytic conversion of $CO_2$ into formic acid with superior performance depends on the accurate understanding of the mechanical process (Figure 2E)(*50-52*). It is generally accepted that, in the initial stage, $CO_2$ is adsorbed on the catalyst surface to form $CO_2$*, where * indicates a binding site at the catalyst surface. The adsorption and desorption of $CO_2$ on the catalyst surface can be precisely regulated by introducing defects and tuning the catalyst structures to improve formic acid production. Duan et al. reported that Bi/CeO$_x$ has excellent $CO_2$ adsorption capacity due to the intrinsically enhanced ability to absorb $CO_2$ with Bi and the inherent heterogeneity of amorphous CeO$_{x(53)}$. The formation of $CO_2$* with a reduced energy barrier in Bi/CeO$_x$ facilitated the subsequent conversion into formic acid products.

In the following step, the formed $CO_2*$ obtains one electron to produce $*CO_2·^-$, which is the rate-determining step of the electrocatalytic conversion of $CO_2$ into $HCOOH/HCOO^-$. The corresponding process faces the problem of a kinetically sluggish and excessive energy barrier. The formed $*CO_2·^-$ is further protonated to generate *OCHO, which can finally be reduced to generate $HCOO^-$ by obtaining one electron. Different reaction pathways may occur due to multiple electron and proton transfer processes. Variable intermediate species, such as *COOH, *CO, *COH, and *CHO, could exist in the reaction system, leading to undesired products, such as CO, $CH_3OH$, $CH_4$, and multi-carbon products (Figure 2F)(*54*). Therefore, promoting the formation of intermediates to obtain formic acid and avoiding the production of C-C bonds are necessary to eradicate the barrier from efficient conversion in practical applications. For instance, pure Sn and Bi usually have an unfavorable binding energy of the intermediate for producing formic acid, which can be solved to form Bi-Sn aerogel(*55*). The synthesized Bi-Sn can inhibit the formation of *COOH, which tends to transform into CO, and suppress the hydrogen evolution reaction with less $H_2$ than pure Bi.

Thus far, great achievements have been made in electrocatalytic $CO_2$ reduction into formic acid(*56*). The selectivity of formic acid production can reach more than 90% through variable heterogeneous catalysts, demonstrating considerable potential for commercialization. Although great progress has been achieved, some challenges still exist and should be overcome to meet the criteria for practical applications. For example, the efficient preparation of formic acid is often achieved under alkaline conditions, which helps to form intermediates that tend to generate formic acid and accelerate electron transfers. In contrast, the produced formic acid often exists in the form of formate in an alkaline environment, leading to the need for further acidification treatment with more energy consumption. Therefore, exploring the efficient synthesis of formic acid under acidic conditions is meaningful, paving the way to an efficient one-step synthesis. Furthermore, the hydrogen evolution reaction should be suppressed due to water as the solvent competing with $CO_2$ to participate in the reduction reaction. Electrocatalytic $CO_2$ reduction and hydrogen evolution reactions involve hydrogen transfer, adsorption, and desorption. By regulating the active sites of the catalyst to inhibit the occurrence of the hydrogen evolution reaction, the yield and selectivity of formic acid can be effectively improved. In addition, the reports on the formation of other carboxylic acids, such as acetic and aromatic carboxylic acid, are still rare due to the difficulty of controlling multiple electron transfers (that tend to have side reactions) and the inertness of C-H in organic reactants. Fortunately, these challenges can be overcome by developing high-efficiency catalysts to accelerate the hydrogen transfer, reducing the activation energy required to form intermediates, *OCHO, and strengthening the ability to activate $CO_2$ and promote the reaction with inert organic reactants.

## 2.3 Thermocatalytic $CO_2$ Conversion Process

Recently, thermocatalytic $CO_2$ conversion into carboxylic acids has been fully developed as an alternative strategy for using the $CO_2$ molecule (Figure 2G)(*57, 58*). It is a relatively practical route to convert inert $CO_2$ compared to the photocatalytic or electrocatalytic path. It has apparent advantages and feasibility in preparing carboxylic acids, such as benzoic, acetylenic, amino, and lactic acids, with higher commercial values and wider application prospects. However, the participation of $CO_2$ in the reaction should usually be carried out under high temperature and pressure due to the nature of its remarkable kinetic and thermodynamic stability(*59*). Thus, higher energy consumption is often required in thermocatalytic $CO_2$ conversion into carboxylic acids, and the harsh reaction conditions could introduce great challenges in ensuring the purity of the products. As a result of recent extensive and in-depth research on the thermocatalytic conversion of $CO_2$ to carboxylic acids, powerful strategies, such as improved atomic utilization and a regulated coordination environment, electronic structure, and space chemical environment of active sites, have been developed to drive the inert $CO_2$ to participate in the reaction under relatively mild conditions.

According to different reactants reacting with $CO_2$, the reaction types of thermocatalytic $CO_2$ into carboxylic acids primarily include the hydrogenation of $CO_2$, Friedel–Crafts -acylation reaction, and carboxylation of terminal alkynes. Hydrogenation of $CO_2$ to produce formic acid has been regarded as a mature and promising route to ensure the effective utilization of $CO_2$ and economic viability(*60*). However, it is not thermodynamically favorable with the free energy change ($\Delta G^0$) of 33.0 kJ mol$^{-1}$ ($CO_2$(g) + $H_2$(g)/HCOOH(l)). The inherent thermodynamic obstacles could be overcome by reacting in an aqueous solution to make the conversion of $CO_2$ into formate more likely(*61*). When the hydrogenation of $CO_2$ occurs in the liquid phase, the reaction ($CO_2$(aq) + $H_2$(aq)/HCOOH(aq)) is more prone to occur with the decline in free energy ($\Delta G^\theta = -4$ kJ mol$^{-1}$). However, the conversion efficiency is still far from satisfactory if no catalyst is involved. The utilization of heterogeneous catalysts has been emphasized to accelerate $CO_2$ conversion by enabling $CO_2$ capture and facilitating the formation of specific intermediates, such as carbonate and bicarbonate, according to the possible mechanism process.

In a typical $CO_2$ hydrogenation process, it is widely accepted that $H_2$ is first dissociated into a hydride species and proton through a heterolytic process (Figure 2H)(*62, 63*). The formed hydride species becomes an intermediate of metal-bound formate by undergoing a nucleophilic attack on the carbon atom of $CO_2$ via metal-hydride bonds, which is usually regarded as the rate-determining step in the $CO_2$ hydrogenation process. Then, formate is formed by reducing the produced intermediate using hydride, instantly recovering the surface of active sites. Although $CO_2$ hydrogenation can produce formic acid, some other challenging issues still exist and should be solved. For instance, organic amine bases, such as ethanolamine, are added to the reaction system to overcome unfavorable free energy, leading to increased costs and difficulty in separation and purification. Furthermore, the reaction of $CO_2$ hydrogenation involves a gas-liquid-solid multi-phase, and this complex reaction system causes obstacles in exploring the mechanisms for the activation and transformation of $H_2$ or $CO_2$ on the surface of selected catalysts. Fortunately, abundant heterogeneous catalysts with different chemical environments and clear structures have been developed to obtain satisfactory performance and deeply investigate the inner origin of catalytic activity. In future studies, more research should explore efficient non-noble heterogeneous catalysts, understand the factors affecting $H_2$ adsorption/activation on non-noble metal sites, and search reasonable methods to tune the microenvironments of the active sites to activate $H_2$ and $CO_2$.

The preparation of multi-carbon carboxylic acid products via carboxylation of terminal alkynes and Friedel–Crafts acylation reactions exhibit outstanding advantages that include a satisfactory yield and green reagents over other synthetic routes, such as the acid-mediated hydrolysis of nitriles(*64*), carbonylation of organic halides with toxic and odorless CO(*65*), and oxidation of primary alcohols (Figure 2I)(*66*). In the carboxylation of terminal alkynes with $CO_2$, unsaturated terminal alkynes are deprotonated using an inorganic base and are coordinated onto the surface of metal sites. Then, the captured electrophilic $CO_2$ is further inserted into the C≡C-metal bond to provide the intermediate, regarded as the rate-determining step for the entire process. Finally, the product, acetylenic acid anion, is released from the catalyst surface. However, heterogeneous catalysts containing Ag, Au, or Cu have been reported to catalyze the conversion of terminal alkynes into carboxylic acids with relatively good activity(*67*). Some difficulties include the need for an additional strong base and the inertness of weakly nucleophilic terminal alkynes that have difficulty attacking $CO_2$. These problems lead to the need for acidification to obtain carboxylic acids, posing a serious challenge to the catalyst stability and poor conversion efficiency. Based on years of exploration, some practical strategies, such as enhancing the ability to capture $CO_2$, controlling the electronic state of metal sites, and optimizing the utilization of active metal centers, have been adopted to overcome existing challenges. It is expected to develop novel heterogeneous catalysts that can activate alkynes and $CO_2$ under acidic conditions and reduce the activation energy of forming acetylenic acid, achieving enhanced efficiency in carboxylic acid production.

The Friedel–Crafts acylation reaction to prepare aromatic carboxylic acids dates to the $Al_2Cl_6$/Al catalytic system created by Olah et al(*68*). The accepted pathway involves an initial complexation between $CO_2$ and metal sites to form a $CO_2$-metal complex(*69*). The produced electrophilic complex reacts with arene, turning into carboxylate coordinated on the catalyst surface and is identified as the rate-determination step. Subsequently, the final product, aromatic carboxylic acids, can be obtained via deprotonation and proton transfer. However, various homogeneous systems, such as $TiCl_4$, $FeCl_3$, $Ga(OTf)_3$, and $CF_3SO_3H$, and heterogeneous systems, including metal nanoparticles (NPs), metal–organic frameworks (MOFs), and others, have been explored to boost the reaction occurrence. They still have limited ability to activate C-H bonds in the benzene ring and inert $CO_2$ and exhibit poor regioselectivity or stereoselectivity for aromatic substrates with the ortho or meta position to participate in the carboxylation reaction. Benefiting from the increased improvement in material characterization methods and increased maturity of preparation methods, the design of heterogeneous catalysts with a controllable spatial environment or surface confinement effect could effectively determine product selectivity by affecting the kinetic behavior, activation energy, and intermediate types.

## 3. Engineering Active Sites to Accelerate $CO_2$ Conversion

Highly active heterogeneous catalysts play a decisive role in converting $CO_2$ to carboxylic acids. Practical strategies can improve the activity, such as maximizing exposed active sites by constructing composite catalysts or tuning the catalyst morphology and increasing the intrinsic activity of the active site via doping heteroatoms and surface functionalization to enhance the activity of traditional catalysts(*70*). These strategies can sometimes work together in the same catalytic system to realize great improvements in activity and selectivity. In the following section, these strategies have been well explained, whereas the corresponding relationships between the structure and catalytic property for each strategy are briefly discussed.

### 3.1 Composite Structure

The composite catalysts primarily refer to the combination of different materials to form supported heterogeneous catalysts, enriching the structure types and delicately enhancing catalytic properties compared with single-component catalysts (Figure 3A)(*71-75*). Specifically, the components containing metal sites, such as metal nitride, metal complexes, oxides, noble metals, and metal clusters, are often loaded onto functional carriers with auxiliary catalytic functions(*76-78*). The reported functional emerging carriers include traditional metal oxide or metal hydroxide and involve emerging support ranging from carbon-based materials (graphene oxide, carbon nanotubes, carbon nitride, and carbon foam) to porous materials (MOFs, COFs, conjugated microporous polymer, and zeolite)(*79-83*).

In the early stages of the development of composite catalysts for accelerating $CO_2$ conversion, traditional carriers, such as metal oxide, are often used to support noble metals or immobilize homogeneous metal complexes(*84, 85*). The catalytic activity of this type of composite catalyst can be further improved by introducing defects, increasing the loading of active species, and other methods. As a result of the weak interaction between the active constituent and carrier, the stability of synthesized composite catalysts should be guaranteed by introducing composites with strong interactions. Recently, based on an elaborate design, a multiple-composite system composed of $Cu_2O$-Pt/SiC/$IrO_x$ proved to be an efficient artificial photosynthetic catalyst for converting $CO_2$ into formic acid due to the greatly prolonged lifetime of photogenerated electrons and holes (Figure 3B-D)(*86*). This study realizes the separate operation of $CO_2$ reduction and $H_2O$ oxidation, facilitating the suppression of the backward reaction of products and promoting whole conversion efficiency.

Another promising composite catalyst to enhance the intrinsic activity of active centers is selecting carbon-based materials as functional carriers. Among these carbon-based materials, graphene, carbon nanotubes, and carbon foam are regarded as appealing carriers due to their good electrical conductivity and nanostructure, conducive to enhancing the material transfer rate(*87, 88*). The inherent activity of the active center is improved by ameliorating the poor electron conductivity and inefficient electron transfer, and the turnover number and turnover frequency of the composite catalysts increase greatly by fully exposing and dispersing the active center. Lou et al. reported that an emerging composite heterogeneous catalyst composed of $Bi_2O_3$ NSs and a conductive multi-channel carbon matrix could catalyze the $CO_2$ reduction into formic acid with high activity (Figure 3E-G)(*89*). This superior performance can be attributed to the synergistic effects of the composite matrix, facilitating electron transfer and enhancing $CO_2$ by producing pyrrolic-N and pyridinic-N. Furthermore, another carbon-based material, carbon nitride, has a good ability to absorb light, and the abundant nitrogen sites are beneficial to capture $CO_2$ and improve the turnover frequency of the active centers(*90*).

Compared with carbon-based composite catalysts, porous material-based composite catalysts have unparalleled advantages due to their ultrahigh specific surface area, adjustable pore environment, and abundant acid or basic sites(*91*). The porous structure can enhance the catalytic efficiency of the active site by capturing $CO_2$ around the active site and increasing the substance transfer rate. Furthermore, the introduced basic sites and Lewis acid sites in the support are beneficial to activate the $CO_2$ in a nonlinear configuration, which promotes the declination of the activation energy required for the rate-determining step of the reaction. In addition, some conjugated porous materials, such as MOFs, with the ability to capture and activate $CO_2$ molecules are used to construct sandwich-shaped composite catalysts to produce alkynoic acid with $CO_2$ (Figure 3H-J)(*92*). Another type of porous material, COFs, can combine active centers to realize the efficient photocatalytic reduction of $CO_2$ to formic acid upon visible-light irradiation due to the excellent visible-light harvesting ability and extended life of photogenerated charge carriers via accelerating the electron transfer(*93*).

**3.2 Heteroatom Doping**

Heteroatom doping is another effective strategy to enhance the activity of heterogeneous catalysts in converting $CO_2$ into carboxylic acids, which can be widely used to form metal-doped oxides, metal alloys, intermetallic compounds, multiple metals with porous materials, and multi-metal complexes (Figure 4A)(*94-100*). In addition, some nonmetal-doped heterogeneous catalysts, including doping nonmetal elements, such as N, P, S, and B, into carbon-based materials or metal-containing materials are designed to boost the catalytic conversion of $CO_2$(*101-103*). Benefiting from the introduction of doped elements, the electronic structure, coordination environment, and geometric configuration of pristine active sites change significantly, leading to enhanced catalytic activity compared with the original state.

Based on the low costs of preparation, nonmetal-doped heterogeneous catalysts have aroused enthusiasm in promoting the preparation of carboxylic acids from $CO_2$. Doping heteroatoms can endow pristine catalysts with improved electrical conductivity and abundant catalytic active centers(*104*). For instance, nonmetal-doped heterogeneous catalysts, such as boron-doped graphene, can catalyze $CO_2$ into formic acid with higher Faradaic efficiency than that of pure graphene(*105*). The enhanced catalytic performance can be attributed to the stronger adsorption of activated $CO_2$ resulting from the high spin density of the introduced boron element. As nonmetal atoms have a weaker activation effect on $CO_2$ than metal atoms, the increase in the inherent activity of the material itself is limited.

Given the limited activity of nonmetal catalytic systems, more attempts have been made to dope heteroatoms on heterogeneous catalysts containing metal atoms(*106-110*). Doping sulfur into indium catalysts has been reported to catalyze the electrocatalytic $CO_2$ reduction to formic acid with excellent catalytic performance due to the strong ability of sulfur to activate water to form hydrogen species, which can further react with $CO_2$ to produce formate (Figure 4B-D)(*111*). Theoretical calculations reveal that doping sulfur into indium significantly decreases the Gibbs free energies (ΔG) for forming HCOO* and HCOOH* in the pathway of forming HCOOH compared with the undoped element. Similarly, doping a heteroatom, such as N, in Sn or Bi containing heterogeneous catalysts can enhance the inherent activity of the initial active sites for the electrocatalytic conversion of $CO_2$ into formic acid. The introduced N element, accompanied by the generation of abundant oxygen vacancies, facilitates the declination of the reaction free energy of HCOO* protonation to form $HCOO^-$ and weakened H* adsorption energy to suppress the hydrogen evolution reaction. In addition, doping heteroatoms into metal oxides can cause configuration distortions, enhancing activity in the hydrogenation of $CO_2$ into formic acid by facilitating the process of $H_2$ dissociation and accelerating the following rate-determining step for forming the HCOO* intermediate with a declined energy barrier compared with the undoped state.

The multi-metal-doped catalytic system is usually regarded as a more effective strategy to tune the inherent activity of the active center than the mentioned strategies due to the significant change in the electronic state and coordination environment, which can have a decisive influence on activating the $CO_2$ and stabilizing intermediates(*55, 112-118*). For instance, doping metals to form alloys has been widely used to boost the hydrogenation of $CO_2$ into formic acid (Figure 4E-G)(*119*). The reaction process is initially provoked by the dissociation of $H_2$ to afford a metal-hydride species and undergoes the formation of the formate intermediate as the rate-determining step for the entire process. Benefiting from doping another metal atom with the electron donation effect, the electron-rich formed metal-hydride species more easily forms formate intermediate through a nucleophilic attack, leading to a decreased energy barrier of the intermediate and corresponding enhanced yields of formic acid(*120*). Usually, Cu-based catalysts produce hydrocarbons in electrochemical $CO_2$ reduction, exhibiting limited selectivity toward a specific product. Doping single-atom Pb to form a $Pb_1Cu$ catalyst can exclusively catalyze $CO_2$ into formate with high selectivity with over 95% Faradaic efficiency(*121*). Theoretical calculations have revealed that the $Pb_1Cu$ electrocatalyst could facilitate forming HCOO* rather than the COOH* path, leading to formate as the main product (Figure 4H-J).

### 3.3 Morphology Control

Tuning the morphology of the heterogeneous catalysts effectively improves the insufficient utilization of catalytic sites of inherent catalysts(*122-124*). In general, the categories of morphology for the designed heterogeneous catalysts can be divided into hollow structures, one-dimensional nanowires, two-dimensional (2D) NSs, quantum dots, and hierarchical porous structures (Figure 5A)(*125-128*). The changes in catalyst morphology have positive effects on the specific surface area, exposed active site, dispersibility in solution, and transfer rate of reactants and can thus endow additional defects, grain boundaries, unsaturated sites, and improved conductivity, which are beneficial for enhancing activity compared with the unprocessed morphology(*129-132*).

The solid bulk nature of conventional heterogeneous catalysts leads to only a small amount of metal sites on the surface participating in the reaction with a low turnover and poor mass transfer rate. In contrast, the catalytic activity of these heterogeneous catalysts can be enhanced via elaborate morphology control. For instance, hollow heterogeneous catalysts, such as $CuInS_2$, have been approved as excellent electrocatalysts to produce formic acid by facilitating the electron transfer and interaction between the reactant and active site and enriching the $CO_2$ local concentration in the void of heterogeneous catalysts(*133*). Furthermore, a hollow photocatalyst can exhibit higher activity in photocatalytic $CO_2$ reduction than that of its bulk counterpart due to the promoted light absorption ability, abundant active sites, and mass transfer channels. In addition, $SnO_2$ quantum wires with grain boundaries have proved to be efficient catalysts to drive the conversion of $CO_2$ into formic acid and exhibit better catalytic activity than $SnO_2$ NPs (Figure 5B-D)(*134*). The enhanced activity could be attributed to more inherent active sites created by the grain boundary-enhanced effect.

The transformation of layered materials into 2D NSs with a single layer or a few layers can have enhanced and novel properties, such as enhanced conductivity and the facile diffusion of photogenerated charge carriers. The low coordinated atomic sites at the edges of NSs and abundant defects caused by morphological changes can afford significantly enhanced catalytic properties. This result is exemplified in work with NSs containing Bi with abundant defects widely used in $CO_2$ electroreduction to formic acid, benefiting from a fast interfacial charge transfer and facile desorption of products(*135*). The theoretical calculation illustrates that the Gibbs free energy of the rate-determining step forming *OCHO in Bi NSs is lower than that of the pristine counterpart, implying that defective active centers in NSs have a stronger ability to stabilize intermediates (Figure 5E-G). Recently, hierarchically mesoporous $SnO_2$ NSs can be obtained by adjusting the hydrophobic chain length of surfactants and exhibits a higher Faraday efficiency in the $CO_2$ electroreduction reaction than mesoporous $SnO_2$ NPs due to the surface mesoscopic structure with increasing surface area, exposure of more active sites, and facile interfacial charge transfer (Figure 5H-J)(*136*).

**3.4 Surface Functionalization**

Surface functionalization can effectively combine the merits of organic and inorganic components and greatly improve the catalytic activity of the inherent catalytic centers for converting $CO_2$ into carboxylic acids (Figure 6A)(*137-139*). The common organic units, including amide, Schiff base, amine, hydroxyl, pyridine, azazole, and light-adsorbing organic ligands, can be flexibly modified using chemical bonds on the surface of typical heterogeneous catalysts to meet the need to improve catalytic activity(*140-142*). Surface functionalization is suitable for a wide range of traditional heterogeneous catalysts, such as metal oxide, metal clusters, metal NPs, porous materials, and others(*143-145*). Moreover, some metal-containing organic molecules, such as porphyrin, phthalocyanine, and the Schiff base complex, can be grafted onto the surface of the original catalysts to boost catalytic performance(*146, 147*).

Functionalization of organic groups on the surface of the active constituent allows many positive facilitators, such as an improved charge transfer rate, strong light-responsive ability, enhanced capture of $CO_2$, and stabilization of intermediates(*148*). For instance, the semiconductor functionalized with an organic linker can obtain higher selectivity in photocatalytic $CO_2$ reduction to formic acid than an untreated one(*149*). The introduced organic constituents could smoothly allow efficient hole and electron migration on the catalyst surface, leading to the enhanced conversion efficiency of $CO_2$. Furthermore, some organic groups with enhanced $CO_2$ capture capacity, such as amide linkages and tris-N-heterocyclic carbene, can be introduced in heterogeneous catalysts to realize the selective electrocatalytic $CO_2$ reduction to formic acid by enhancing the concentration of $CO_2$ around the active site and suppressing the formation of intermediates to produce CO (Figure 6B-D)(*150*). The theoretical calculations have found that the tris-N-heterocyclic carbene group on the surface of Pd favored the formation of COOH* and the hydrogenation of COOH* into HCOOH with a lower energy barrier than that of the pure Pd catalyst.

Recent studies have noted the importance of surface functionalization in enhancing the stability of the original heterogeneous catalyst and auxiliarily activating the reaction substrate. For instance, a bipyridine-based covalent triazine framework with an Ru active center has been reported in catalyzing $CO_2$ hydrogenation with high activity. However, the bond between the metal and bipyridine is not strong enough, and the active center could dissociate into the reaction solvent, leading to poor stability of the designed catalyst (Figure 6E-G)(*151, 152*). An oxyanionic ligand is a well-established approach in stabilizing the active species by avoiding the dissociation of metal sites from bipyridine. Moreover, the introduced oxyanionic ligand could assist in the heterolysis of $H_2$, which is the rate-determining step for this conversion, leading to highly efficient conversion.

Organic functional groups can form specific intermediates to promote the reaction rate due to the reduced activation energy of the rate-determining step. For instance, different organic amines can be flexibly grafted onto the surface of noble metals, such as Au NPs, to catalyze $CO_2$ hydrogenation, which can stabilize noble metals to avoid aggregation and enhance $CO_2$

concentration(*153*). More importantly, these grafted organic amines can help activate $CO_2$ through a nonbicarbonate route with negative free energies for adsorption of $CO_2$ and can benefit from forming the zwitterion intermediate, which is regarded as beneficial for producing formic acid (Figure 6H-J). The density functional theory (DFT) calculation has revealed that the added Schiff base plays a vital role in the activated $CO_2$, which can be further hydrogenated by the activated H to form an $HCO_2$ intermediate and become HCOOH with the assistance of another H atom.

## 4. Advanced Characterization Techniques and Theoretical Calculations

Although great progress has been achieved in the catalytic conversion of $CO_2$ into carboxylic acids, previous studies have suffered from a lack of clarity in revealing the dynamic process of the $CO_2$ transformation. Considering that the catalytic reaction and product selectivity are dominated by the evolution of the catalyst structure in the reaction process and the interaction between the active sites and reactants or intermediates, it is crucial to distinguish the true active centers and their dynamic evolution in the reaction process(*154*). However, identifying real active sites and monitoring the evolution of the catalysts or involved reactants has been vigorously challenged by dynamic changes in the catalyst structure during the reaction(*155*). In addition, difficulties exist in distinguishing complex intermediates due to multi-step reaction paths in converting $CO_2$ into carboxylic acids(*156*). Thus, these obstacles preclude the recognition of real active sites and reveal the reaction mechanisms of $CO_2$ conversion. As a result of developing numerous emerging in situ characterization instruments, we can approach observing the true state of the catalysts in the reaction process via operando techniques at the molecular level.

Parallel to the characterization technique development, the theoretical calculation should improve the fundamental understanding of the mechanical process in converting $CO_2$ to carboxylic acids. The ongoing development of the DFT has led to significant progress in preparing the catalyst from a blind attempt to accurately predict using computer-aided prediction. Importantly, the reactivity origin of the designed heterogeneous catalysts could be uncovered by analyzing the effect of changes in the electronic state in metal sites, exposed crystal faces, and coordination environment based on DFT calculations. The following section introduces advanced in situ techniques and significant progress in theoretical calculations in converting $CO_2$ into carboxylic acids.

**4.1 In situ Characterization Techniques**

Moreover, X-ray absorption spectroscopy (XAS) has become an essential characterization technique to detect the structural evolution of catalysts during $CO_2$-involved reactions (Figure 7A)(*157*). The specific elements, catalyst coordination environment, and chemical valence changes that affect the conversion efficiency of the $CO_2$ and product selectivity can be determined using XAS (Figure 7B). In addition, XAS can be divided into the extended x-ray absorption fine structure (EXAFS) and X-ray absorption near-edge structure (XANES) spectra based on different origins. The XANES spectrum provides fingerprint information on the geometric structure, oxidation state, and electronic structure of the detected metal sites. Furthermore, the EXAFS spectrum identifies the atomistic structure/configuration and coordination geometry of reactive centers combined with a fitting analysis. Recently, Li et al. have analyzed the EXAFS results by building reasonable models, revealing abundant defects in reduced $Bi_2O_3$ in the $CO_2$ reduction process through a model-based quantitative

analysis(*158*). In addition, the authors have testified that the changes in chemical valences in bismuth could affect the electrocatalytic reduction of $CO_2$ into formate with in situ XAS (Figure 7C).

Compared to XAS, the X-ray photoelectron spectroscopy (XPS) technique could reveal the chemical composition, oxidation states, and electron transfer of designed catalysts with the characteristics of surface sensitivity (Figure 7D). With the development of reaction cells and simplification of operating conditions, the in situ ambient pressure XPS (AP-XPS) technique has been exploited to monitor the dynamic changes in the electronic state and structural evolution in the reaction process (Figure 7E). For instance, Yoshinobu et al. have investigated the reaction process of $CO_2$ hydrogenation with Zn/Cu alloy using AP-XPS, which reveals that the $CO_2$ can be activated on the surface of the Zn/Cu to form the carbonate species under reaction conditions (Figure 7F)(*159*).

In situ Raman spectroscopy is another valid characterization technology that complements IR spectroscopy to detect catalyst evolution and possible intermediates in a liquid solution (Figure 7G). The intrinsic vibration and rotation energy levels of catalysts in the Raman spectrum can reflect their intrinsic properties and are expected to probe the mechanism process for the corresponding catalytic systems (Figure 7H). Song et al. have used in situ Raman spectroscopy to probe the transition of intermediate adsorption states in electrochemical $CO_2$ reduction over S-doped Cu-based catalysts. The intensity of peak at 1080 cm$^{-1}$ corresponding to the symmetric stretching adsorption of $CO_3^{2-}$ is more obvious over the S-doped Cu-based catalyst than the undoped one, manifesting that S doping can enhance the adsorption ability of carbonate intermediates (Figure 7I)(*160*).

Compared with the above-mentioned techniques, in situ IR spectroscopy has advantages in detecting the intermediates formed on active sites during the reaction, benefiting from its high sensitivity to the vibration mode of the adsorbed molecules (Figure 7J). The IR absorption peak position is the characteristic signal of the adsorbed molecule, reflecting the structural composition and chemical groups in the reaction process (Figure 7K). For example, Gong et al. have investigated the active species in $CO_2$ reduction over $SnO_X$ as a catalyst using in situ attenuated total reflection surface-enhanced IR absorption spectroscopy(*161*). The signals of the HCOO* can be identified by analyzing the characteristic peaks when the potential is applied to the surface of Sn-OH branches. In addition, the intensity of HCOO* can be quantitatively analyzed to evaluate the content of produced intermediates, assessing the influence of different hydroxyl content on catalytic activity (Figure 7L).

**4.2 Theoretical Calculations**

Extraordinary progress in theoretical calculation has created enormous opportunities in evaluating the energy required for the adsorption, activation, and conversion of reactants in the catalytic reaction process, allowing the design of highly efficient heterogeneous catalysts(*162-165*). For instance, by analyzing the adsorption free energy of $CO_2$ and other reactants on the surface of various catalysts, we can determine whether the adsorption and desorption behavior of reactant molecules affects the conversion efficiency of the entire reaction process and preliminarily assess the catalytic performance of the designed catalysts. Zhang et al. have used theoretical calculations to reveal that the grafted amide bond on the graphene oxide substrate with a lower $CO_2$ adsorption energy of -6.94 kcal/mol demonstrates a stronger $CO_2$ capacity than the original graphene oxide with the $CO_2$ adsorption energy of -4.56 kcal/mol(*166*).

Intermediates are often difficult to capture during experimentation; thus, theoretical calculations introduce substantial possibilities for identifying true intermediates and the optimal configuration on the catalyst surface, which provide essential information for disclosing the entire reaction paths. For instance, Johnson et al. have reported that frustrated Lewis pairs functionalized MOFs (UiO-66-P-BF$_2$) can effectively catalyze the hydrogenation of CO$_2$ (Figure 8A-B)(*167*). They further identified the desirable reaction path for CO$_2$ hydrogenation over UiO-66-P-BF$_2$ by comparing the energy barrier of two possible reaction pathways. The corresponding theoretical calculation results demonstrated that the reaction between physically-adsorbed CO$_2$ and chemisorbed 2H* with an energy barrier of 0.47 eV is more favorable than the process of reacting between physically-adsorbed H$_2$ and CO$_2$* with an evident higher barrier of 2.65 eV, verifying the possible mechanism proceeded by H$_2$ heterolytic dissociation, followed by the CO$_2$ reaction with the adsorbed H atoms (Figure 8C).

Moreover, the rate-determining step in the corresponding process can be distinguished by comparing the energy barrier in the formation and transfer of reactants, the intermediate state, and products on the catalyst surface. For instance, theoretical calculations have revealed that photocatalytic conversion of CO$_2$ into formic acid over COF-367-Co with a tunable spin state undergoes two primary transition states(*168*). These states form an O-H bond between the oxygen in CO$_2$ and the hydrogen with an energy barrier of 0.13 eV and the formation of HCOOH* via another transition state with an energy barrier of 0.68 eV (Figure 8D-F). The latter step is regarded as the rate-determining step for the overall reaction. By comparing the energy of the rate-determining step of different spin-state COF-367-Co catalysts, the spin-state transition of Co can be disclosed as the decisive factor in forming formic acid with CO$_2$.

Furthermore, product selectivity in converting $CO_2$ can be explained by evaluating the energy superiority of various products for the designed catalysts. For example, the process of electrocatalytic conversion of $CO_2$ into formic acid is often accompanied by producing CO. By comparing the Gibbs free energies of electrochemical $CO_2$ reduction into the corresponding products, the generation of CO via a CO* intermediate or production of HCOOH(g) via the COOH* intermediate can be predicted over synthesized catalysts(*169*). In addition to explaining the fundamental reasons for product selectivity, the occurrence of competing reactions on the surface of catalysts can be inhibited by optimizing the catalyst structure.

With the development of machine learning and theoretical calculation methods, more efficient heterogeneous catalysts are predicted and synthesized to boost the conversion of $CO_2$-involved reactions(*170*). For instance, Guo et al. have successfully used machine learning to electro-catalyze $CO_2$ into formic acid with high yield by discovering and optimizing additives in Cu-based catalysts (Figure 8G)(*171*). In the initial step, they select combinations from the collected additive library, including different metal salts and water-soluble organic molecules, to tune the morphology and surface structure of Cu catalysts (Figure 8H). Then, through the screening of machine learning, the experiment is guided using Sn containing metal salts and organic ligands containing aliphatic amino and carboxyl groups to obtain high selectivity of formic acid ($FE_{HCOOH}$ = 65%) (Figure 8I). In the future, machine learning is expected to play a more important role in achieving efficient conversion and preparation of a variety of high-value carboxylic acids.

## 5. Frontier Progress of Conversion of $CO_2$ into Carboxylic Acid

### 5.1 Photoreduction of $CO_2$ into Carboxylic Acid

The synthesis of multi-carbon carboxylic acids with a high conversion rate is challenging due to the transfer of more electrons involved in the photoelectric conversion process. Some breakthroughs have been made in preparing multi-carbon carboxylic acids, such as acetic acid, through photocatalytic $CO_2$ reduction(*172*). Wang et al. have realized the efficient preparation of acetic acid by constructing a hybrid photosynthetic system comprising semiconductors and bacteria(*173*). Perylene diimide derivative (PDI) and poly(fluorene-co-phenylene) (PFP) are selected to coat the bacterial surface as photosensitizers to form a p-n heterojunction layer, enhancing the efficiency of the hole/electron separation (Figure 9A). The designed conjugated molecules/bacteria could exhibit a higher amount of the acetic acid under illumination than only bacteria in the light or conjugated molecules/bacteria in the dark. The acetic acid yield for PDI/PFP/bacteria and PDI/bacteria is 0.63 and 0.25 mM, respectively, which has great potential to further achieve higher activity by optimizing the structure and composition (Figure 9B-C). This work presents the advantages of combining biological bacteria and semiconductors in photocatalytic carboxylic acid production.

Furthermore, Sun et al. have developed an artificial photocatalytic system to produce acetic acid by constructing abundant exposed surface oxygen vacancies in ultrathin $WO_3 \cdot 0.33H_2O$ nanotubes (Figure 9D)(*174*). The acetic acid yield over synthesized $WO_3 \cdot 0.33H_2O$ with high oxygen vacancies can reach 94 μmol/g after 10 h and is much higher than samples with low oxygen vacancies and commercial $WO_3$, testifying the important role of oxygen vacancies in the $CO_2$ reduction process (Figure 9E). In situ diffuse reflectance IR Fourier-transform spectroscopy has been further used to explore the role of oxygen vacancies in promoting the conversion of $CO_2$ into acetic acid. The results reveal that the $HCO_3$ species is an important intermediate for producing acetic acid, which can be transformed into the COOH intermediate under light irradiation. Specifically, the C=O bond of the $CO_2$ is activated by the W-OH group in $WO_3$ upon light irradiation and is transferred into the O=Ċ-OH radical intermediate after obtaining photogenerated electrons. The corresponding $HCO_3$ bicarbonate species can be formed at the place of oxygen vacancies by undergoing a proton-coupled electron transfer and further reaction with the adjacent ·COOH radicals to form C-C bonds by departing from the oxygen vacancy. After a few steps of proton-coupled electron transfer under light irradiation, acetic acids have been produced with a high yield (Figure 9F).

In addition, a great breakthrough has been made in promoting $CO_2$ to participate in the photocatalytic organic reaction to produce carboxylic acids (Figure 9G). Some organic carboxylic acids with important application values, such as trans-cinnamic acid and diaryl α-amino acids, can be obtained under blue-light excitation or sunlight. For instance, Schmalzbauer et al. have reported that a redox-neutral C-H carboxylation of arenes and styrenes can be realized by reacting with $CO_2$ to afford carboxylic acids using an anthrolate anion photocatalyst under the blue-light excitation (Figure 9H)(*175*). Then, they have proposed a possible reaction mechanism for the photocatalytic C-H carboxylation of the heterocycle. In the initial step, the catalyst, 2,3,6,7-tetramethoxyanthracen-9(10H)-one (TMAH), is deprotonated to form an anionic species with the aid of a base. Next, the excited anion TMA* is formed under visible light (455 nm) and is quenched by the organic substrate (arene) to produce the radical TMA and radical arene anion. Afterward, $CO_2$ attacks the radical arene anion to generate a radical carboxylate intermediate that is deprotonated immediately to produce a carboxylic anion in the presence of the base (Figure 9I). The final products are obtained by further treatment with hydrochloric acid. This work presents new opportunities for an atom-economic and energy-efficient utilization of $CO_2$ in producing aromatic carboxylic acids and drug intermediates.

**5.2 Electroreduction of $CO_2$ into Carboxylic Acid**

Selective production of formic acid from $CO_2$ electroreduction over abundant catalysts has achieved ideal activity of over 90% Faradaic efficiency. However, the neglected challenge concerning mixed products in an alkaline aqueous system should be overcome to satisfy the practical applications of electrocatalytic preparation of renewable formic acid. The generated formic acid in the conventional H-cell or flow-cell reactors is formate ions mixed with other ion impurities, such as $K^+$ and $HCO_3^-$. This problem can be overcome by improving the reaction device. For instance, Xia et al. have developed an efficient strategy to resolve this challenge by decoupling the ionic conduction and allowing product collection(*176*). Specifically, a porous solid electrolyte (PSE) layer is used to replace the traditional liquid electrolytes, promoting fast ionic transportation and recombining the generated formate and proton to form pure formic acid (Figure 10A). The selected insoluble solid electrolyte provides a proper platform to collect the formic acid efficiently by flowing the deionized water through the PSE layer. Furthermore, they have synthesized an ultrathin 2D Bi catalyst to catalyze $CO_2$ into formic acid using the new $H^+$-conducting solid electrolytes. A high $FE_{HCOOH}$ of 93.1% is achieved, and negligible amounts of other ion impurities have been detected using inductively coupled plasma atomic emission spectroscopy (Figure 10B).

Although it has been demonstrated that pure formic acid can be obtained with solid electrolytes, the formic acid concentration is limited due to the need for a significant amount of water. Recently, Fan et al. have realized continuous generation of high-purity and high-concentration formic acid using an all-solid-state electrochemical device in which the generated formic acid can be efficiently collected via an $N_2$ stream flowing through the PSE layer (Figure 10C)(*177*). High activity (maximal Faradaic efficiency ~97%) and ultrahigh concentrations of pure formic acid solutions (up to nearly 100%) can be obtained over the Bi catalyst with an abundant grain boundary (Figure 10D). This emerging system is expected to be suitable for modular and high-pressure systems, which have exciting potential in future large-scale production of formic acid. Future research should develop highly stable and active catalysts and solid electrolytes with excellent performance and efficient ion-exchange membranes to realize low energy consumption to convert $CO_2$ into formic acid.

In addition to the preparation of high-concentration pure formic acid, there are numerous attempts for further progress in producing multi-carbon carboxylic acids via the electrocatalysis of $CO_2$(*178, 179*). For instance, Genovese et al. have reported that the selectivity to acetic acid (61%) could be reached in the three-electrode cell over nanostructured ferrihydrite-like (Fh-FeOOH)/nitrogen-doped carbon (Figure 10E)(*180*). This high activity for acetic acid production can be attributed to the formed nitrogen-coordinated iron (II) sites at the interface between iron oxyhydroxide and the support, demonstrated by operando x-ray spectroscopy techniques (Figure 10F). Compared to producing formic acid from $CO_2$, the selectivity is far from satisfactory due to possible competitive reactions. More effort should be made to design effective catalysts to enable C-C coupling.

## 5.3 Thermocatalytic Conversion of $CO_2$ into Carboxylic Acid

In the thermocatalytic conversion of $CO_2$ into carboxylic acids, the carboxylic acid yield should be improved, and the product selectivity, including regioselectivity, stereoselectivity, and chemoselectivity, should also be controlled by the desired catalysts. Great achievements have been made in recent years, profiting from long-term efforts and exploration(*181*). For instance, Martin et al. have reported that a host of unactivated alkyl chlorides can react with $CO_2$ to produce carboxylic acids with a high conversion rate over catalysts containing Ni (Figure 11A)(*182*). Moreover, an anti-carbometalation product has obtained with high selectivity based on this designed catalytic system. The high selectivity of the anti-product in converting secondary alkyl bromides may be attributed to forming rapid isomerization of vinyl radical species, which would undergo recombination to produce a different configuration of the Ni(I)BrLn species (Figure 11B). This work realizes exquisite chemoselectivity and a high conversion rate for carboxylic acid production with unactivated alkyl chlorides and $CO_2$.

Ling et al. have exploited the interface of Ag@MOF to provide a pseudo microenvironment with high $CO_2$ pressure to enhance the direct C-H carboxylation at ambient conditions (Figure 11C)(*183*). Specifically, the interface of Ag@MOF comprised an array of Ag nanocubes has been grafted with mercaptophenol as reaction substrates. XPS has been used to characterize the interaction between the substrate and MOFs, confirming the existence of Zn-O bonds between phenoxide and MOF, which cannot be found in methylbenzenethiol functionalized Ag nanocubes without MOF (Figure 11D). As a result of the activation and confinement effects of MOF on reaction substrates, the direct C-H carboxylation of arene occurs at the NP@MOF interface under ambient pressure at 25°C. More importantly, an unprecedented meta-carboxylated arene is generated rather than a traditional ortho-carboxylated product (Figure 11E). This work introduces new opportunities to synthesize high-value products that cannot be obtained under common conditions and provides new inspiration for improving the selectivity of carboxylic acids.

Recently, Xiong et al. have reported that the hydrocarboxylation of alkynes with $CO_2$ can be catalyzed to form a wide array of α-acrylic acids with high regioselectivity through the combination of $Pd(PPh_3)_4$ and 2,2′-bis(diphenylphosphino)-1,1′-binaphthalene (binap) under mild conditions(*184*). For example, for phenylacetylene, the product of α-acrylic acids can be obtained with a 70% isolated yield in high selectivity by adding 0.5 mol% $Pd(PPh_3)_4$ and 0.5 mol% binap (Figure 11F). DFT studies have further revealed that alkynes react with $CO_2$ via the cyclopalladation process, generating a five-membered palladalactone intermediate, and undergo a σ bond metathesis step regarded as the rate-determining step of the entire process. The calculated results have demonstrated that the Markovnikov adducts, α-acrylic acids, are the main products with a more favorable free-energy barrier than the anti-Markovnikov product (Figure 11G). This work reveals that the designed catalyst is a powerful tool to use $CO_2$ to produce complex molecules with high selectivity.

## 6. Conclusions and Prospects

In the past few decades, the exhaustion of fossil fuels and the deterioration of the environment have necessitated carbon neutrality, and prior studies have noted the importance of consuming the emission of $CO_2$ via a stainable and green approach. Accordingly, the direct conversion of $CO_2$ into chemical compounds accomplishes both goals. Climate warming can be slowed, and fine chemicals can be produced in an economically and environmentally friendly way. Considering the wide application prospects and social needs concerning carboxylic acids, the direct conversion of $CO_2$ into carboxylic acids has attracted considerable attention in fundamental research and industrial applications. Given the importance and social needs regarding this field, a systematic and in-depth review of the selective conversion of $CO_2$ into carboxylic acids is urgently needed.

The process of $CO_2$ conversion requires a controllable catalytic system to accelerate the conversion of reactants; thus, recent developments and feasible tuning strategies on designing heterogeneous catalysts are systematically summarized in this review. The past few years have witnessed numerous breakthroughs involving the conversion of $CO_2$ into carboxylic acids, benefiting from the advancements in characterization techniques and the understanding of the mechanical process. However, many issues should still be addressed to achieve large-scale production of carboxylic acid compounds with a high conversion rate and satisfactory selectivity.

First, the capture or identification of intermediates and transition states in the reaction process is at the center of understanding the mechanism of conversion efficiency of $CO_2$ into carboxylic acids. Ultra-short lifetimes of transition states and the instability of the intermediate under characterization conditions have accentuated the problem of the lack of in-depth investigation of intermediates at the molecular level. An appropriate transition state can only be inferred indirectly from some seemingly reasonable in situ characterization techniques. An effective solution is achieving catalytic conversion of $CO_2$ under relatively mild or atmospheric conditions via the design of high-efficiency catalysts, allowing the accurate detection of the transition state of the reaction with the existing in situ characterization techniques. In addition, advanced in situ characterization techniques and a combination of multiple in situ technologies should be strongly developed to track the evolution of reactants, including C-O bond breakage, the formation of C-H or C-C bonds, and possible intermediates.

Second, although remarkable progress has been made in photo-catalyzing $CO_2$ into carboxylic acids, several challenging issues should be considered and solved. The use of near IR light and visible light is far from satisfactory, and low quantum efficiency hinders the widespread practical application of photocatalysis. Increasing the efficiency of photocatalytic $CO_2$ conversion and making the most of natural light by constructing stable heterogeneous catalysts with sufficient activity and a strong ability to harvest broad-spectrum light are highly desired. In addition, the introduction of sacrificial agents or cocatalysts to guarantee conversion efficiency may create additional problems, such as high production costs and poor catalyst stability. The appealing solution is to develop heterogeneous catalysts with multiple functions to improve light utilization and produce synergistic effects between the active sites and cocatalysts and thus to get rid of the usage of sacrificial agents.

Third, electrocatalytic conversion of $CO_2$ into high-value-added products, such as formic acid, is expected to reach the goal of actual production based on the continued development of efficient catalysts and facile reactors. However, the generation of multi-carbon carboxylic acids with high selectivity still has a long way to go due to the competition of multiple side reaction paths. Regulating the active catalytic sites via the electronic state, coordination environment, and chemical components is feasible to inhibit undesired products. Furthermore, the research on disclosing the reconstruction of the catalyst in the electrolyte process is still in its infancy, leading to a poor understanding of the relationship between structure and activity. More attention should be focused on revealing the changes in composition and structure that have occurred in the reaction process, reducing the over-potential required for $CO_2$ reduction and inhibiting the occurrence of competitive reactions under the premise of ensuring the efficiency of carboxylic acid products.

Fourth, the conversion efficiency of the thermo-catalytic conversion of $CO_2$ can be improved under harsh reaction conditions, such as high temperature and high pressure, but introduces substantial challenges to the selectivity of products due to the byproducts caused by the high energy. Especially for carboxylic acid compounds involving organic substrates, the regioselectivity, stereoselectivity, and chemoselectivity of products determine the purity and produce specific chemicals with different values and applications. Highly selective carboxylic acids can be synthesized under relatively mild conditions only by accurately constructing a steric hindrance environment, using the pore restriction effect and regulating the strength of the Lewis acidity and alkalinity.

Fifth, relatively high energy consumption with high temperature and pressure is usually required to obtain satisfactory yields of carboxylic acids, especially for preparing multi-carbon carboxylic acids. The accompanying process poses great challenges to the reaction devices and the stability of catalysts. More efficient strategies are urgently needed to promote the conversion of $CO_2$ under green, environmentally friendly, and mild conditions. It has been demonstrated that seeking strategies to realize photoelectric conversion or electrothermal conversion could improve the efficiency of the input energy and create new opportunities to reduce carbon emissions and significantly decrease production costs.

Sixth, theoretical algorithms play a vital role in the deeper understanding of the reaction mechanisms and correlations between catalytic performance and structure. However, the inherent limitations of theoretical models and the narrow applicability of theoretical methods greatly compromise the guiding role of theoretical calculation. Machine learning and artificial intelligence are highly desirable for the design of experiments, optimization of catalysts, understanding of the mechanism. By analyzing the available fundamental data and basic theory with machine learning, the discovery of efficient catalysts can be accelerated with an accurate guide rather than tedious trial-and-error investigations.

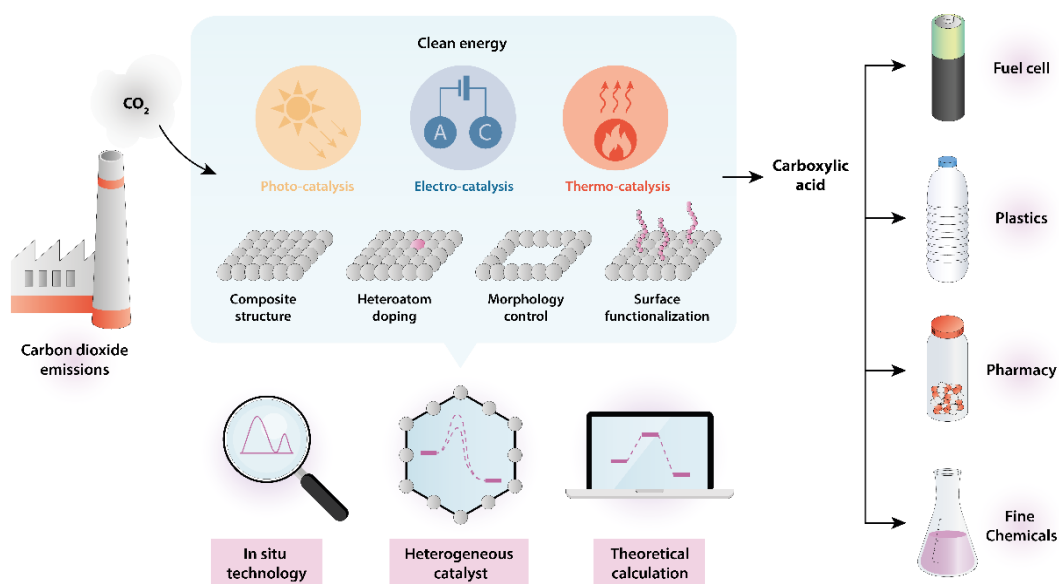

**Figure 1.** Schematic diagram of conversion of carbon dioxide into carboxylic acids.

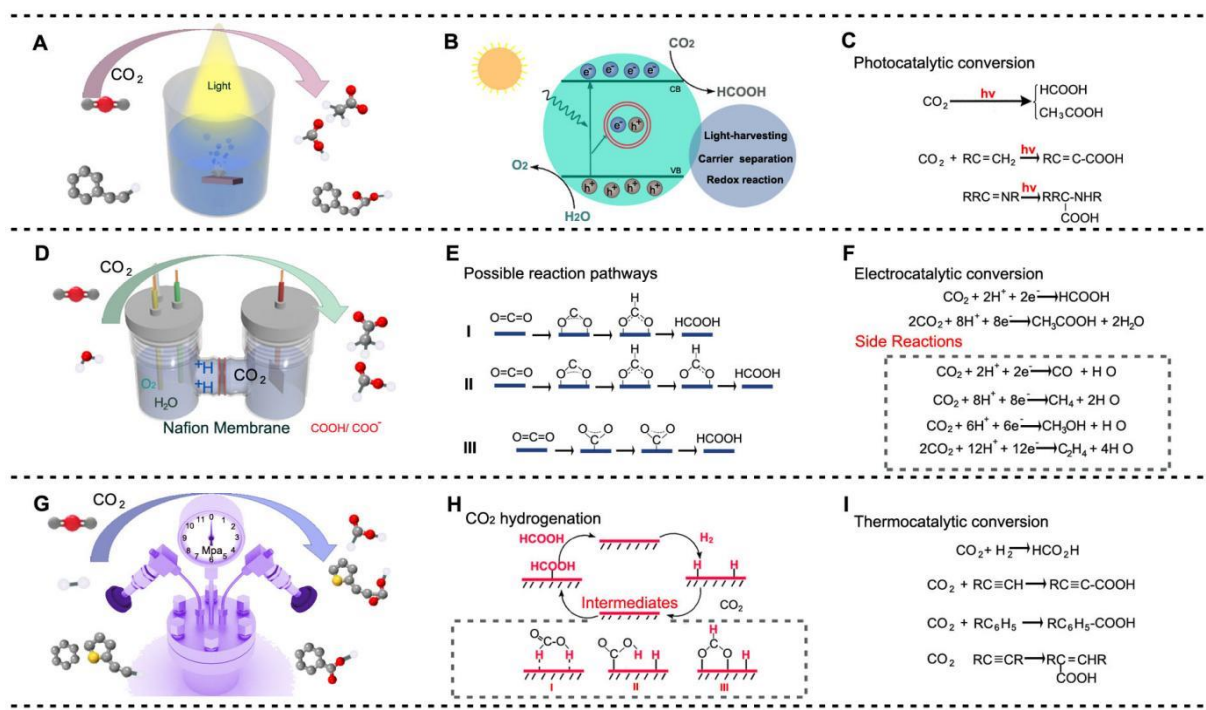

**Figure 2. Conversion of CO$_2$ into carboxylic acids under the drivers of light, electricity, and heat.** (A-C) Photocatalytic conversion of carbon dioxide (CO$_2$) into carboxylic acids: (A) schematic diagram of photocatalytic conversion of carbon dioxide (CO$_2$) into carboxylic acids, (B) mechanism of photo-catalyzing CO$_2$ into formic acid, (C) reaction paths of the photocatalytic process for producing different carboxylic acids. (D-F) Electrocatalytic conversion of CO$_2$ into carboxylic acids: (D) schematic diagram of electrocatalytic conversion of CO$_2$ into carboxylic acids, (E) possible mechanism of electro-catalyzing CO$_2$ into formic acid, (F) reaction paths of the electrocatalytic process for producing different chemicals. (G-I) Thermo-catalytic conversion of CO$_2$ into carboxylic acids: (G) Schematic diagram of the thermo-catalytic conversion of CO$_2$ into carboxylic acids, (H) possible mechanism of hydrogenation of CO$_2$ into formic acid, (I) reaction paths of the thermo-catalytic process for producing carboxylic acids.

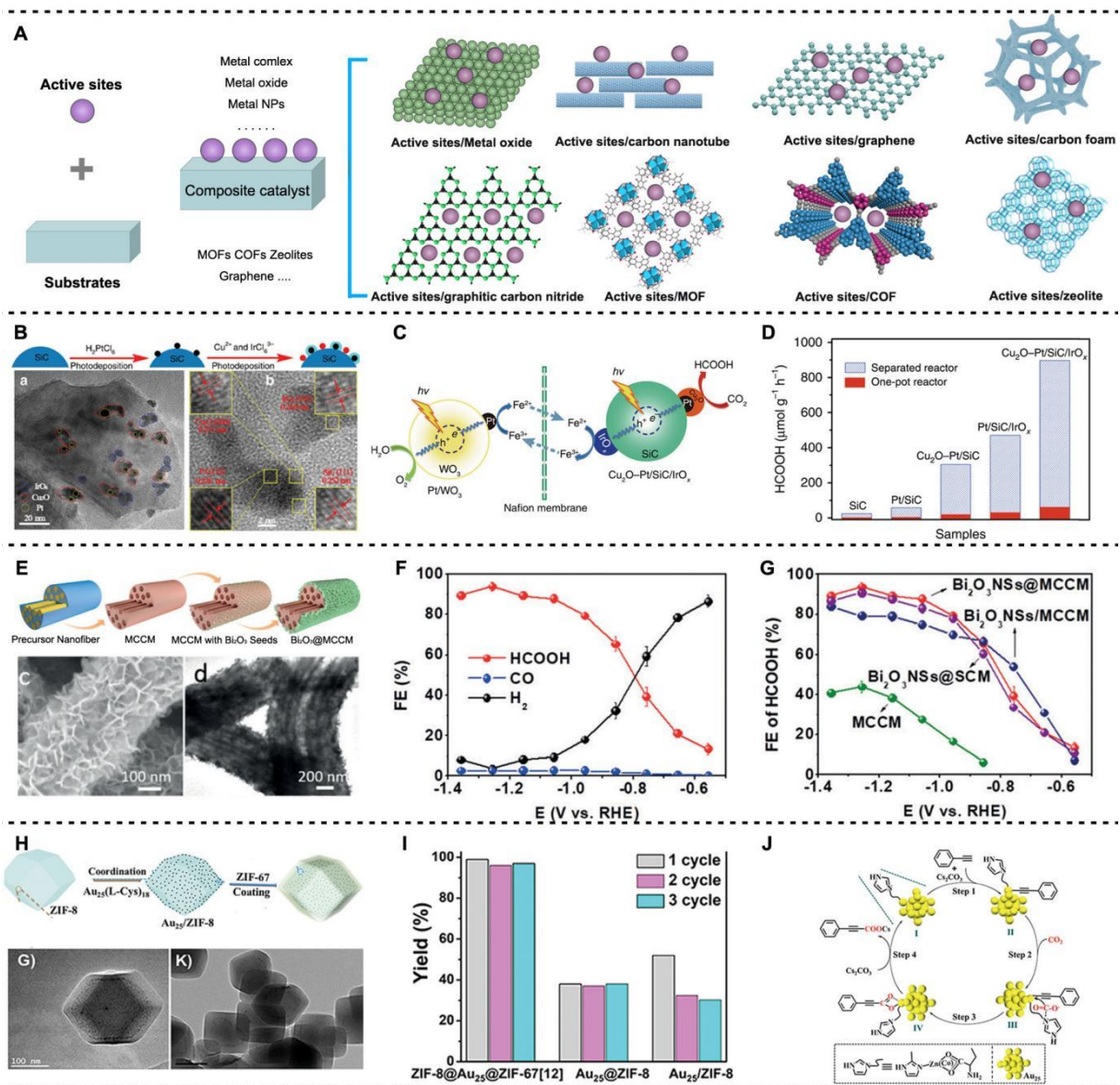

**Figure 3. Conversion of CO$_2$ into carboxylic acids by composite structure.** (A) Schematic diagram of loading active sites onto different supports to form the composite catalyst. (B-D) Photocatalytic conversion of carbon dioxide (CO$_2$) into formic acid by Cu$_2$O-Pt/SiC/IrO$_x$ catalyst: (B) Schematic diagram and TEM images of the Cu$_2$O-Pt/SiC/IrO$_x$ catalyst, (C) schematic diagram of the efficient CO$_2$ reduction and O$_2$ evolution mechanism, (D) HCOOH evolution with Cu$_2$O-Pt/SiC/IrO$_x$ as the photocatalyst [(A-C), adapted with permission from Wang et al. (*86*)]. (E-G) Electrocatalytic conversion of carbon dioxide (CO$_2$) into formic acid by Bi$_2$O$_3$ nanosheets (NSs)/multichannel carbon matrix (Bi$_2$O$_3$@MCCM): (E) Schematic

diagram of Bi$_2$O$_3$@MCCM. (F) FE of all products over the Bi$_2$O$_3$NSs@MCCM. (G) FE of HCOOH of different catalysts [(E-G), adapted with permission from Liu et al. (*89*)]. (H-J) Thermo-catalytic conversion of carbon dioxide (CO$_2$) into acetylenic acid by ZIF-8@Au$_{25}$@ZIF-67[tkn] and ZIF-8@Au$_{25}$@ZIF-8[tkn] [tkn = Thickness of Shell]: (H) Synthetic route for the sandwich structures and TEM images of ZIF-8@Au$_{25}$@ZIF-67[tkn] and ZIF-8@Au$_{25}$@ZIF-8[tkn], (I) catalytic activity of various catalysts for the carboxylation of phenylacetylene and the stability of various catalysts for the carboxylation of phenylacetylene, (J) proposed catalytic mechanism of the reaction between terminal alkynes and CO$_2$ over ZIF-8@Au$_{25}$@ZIF-67[tkn] [(H-J), adapted with permission from Yun et al. (*92*)].

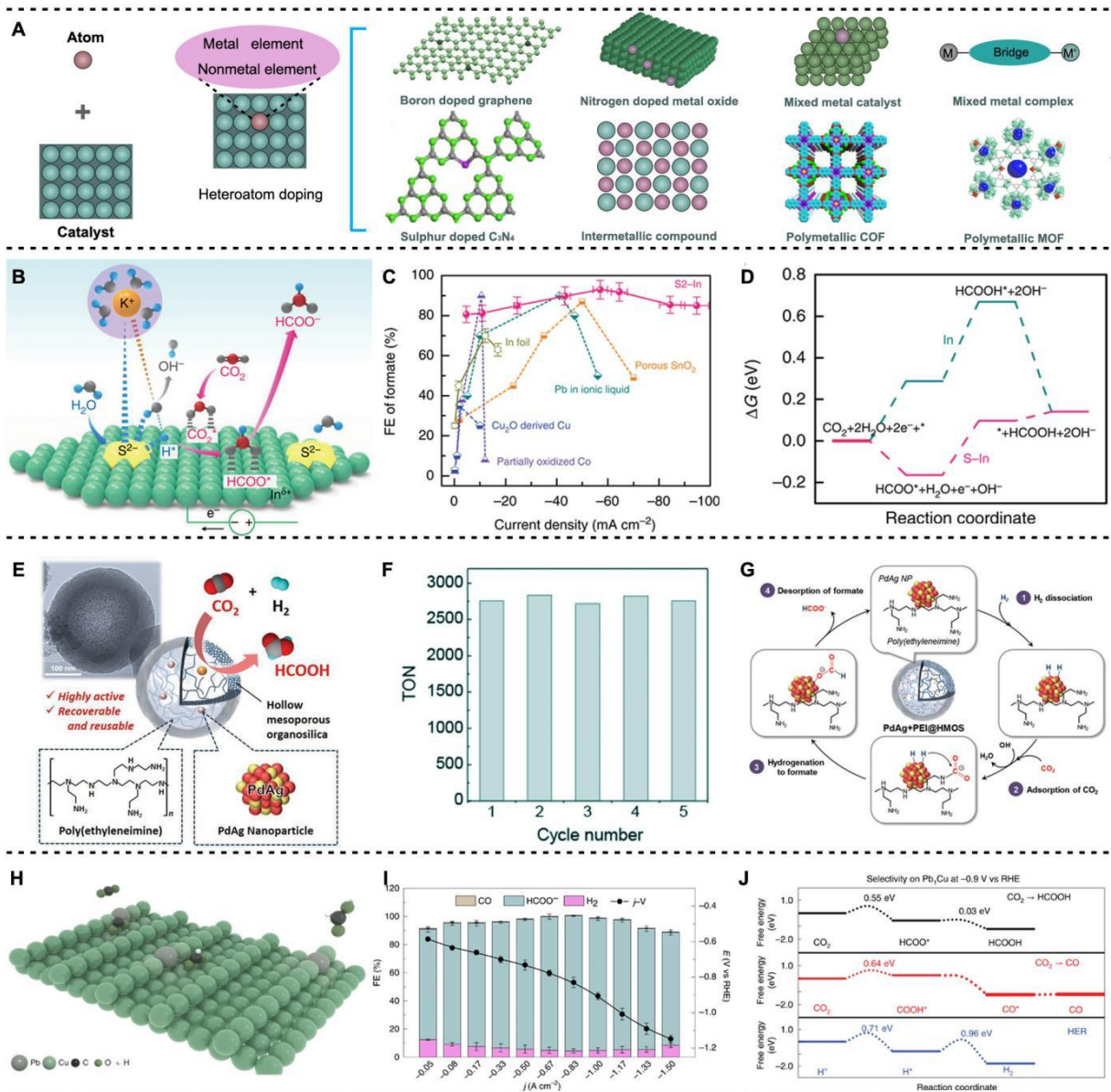

**Figure 4. Conversion of CO$_2$ into carboxylic acids by doping heteroatom.** A) Schematic diagram of forming the doped catalyst. (B-D) Electrocatalytic conversion of carbon dioxide (CO$_2$) into formate by sulfur-doped indium catalysts: (B) Schematic illustration of the role of S$^{2-}$ in promoting water dissociation and H* formation for the reduction of CO$_2$ to formate, (C) plot of FE of formate vs. the current density for the S$^{2-}$In catalyst and some typically reported catalysts, (D) Gibbs free-energy diagrams for CO$_2$RR to HCOOH on S-In (101) and S-In (101) surfaces [(B-D), adapted with permission from Ma et al. (*111*)]. (E-G) Thermocatalytic conversion of carbon dioxide (CO$_2$) into formic acid by PdAg+PEI@HMOS catalyst: (E)

Schematic illustration of $CO_2$ hydrogenation to produce formate with the PdAg+PEI@HMOS catalyst, (F) reusability tests of PdAg+PEI@HMOS for $CO_2$ hydrogenation, (G) plausible reaction mechanism for the $CO_2$ hydrogenation to produce formate over PdAg+PEI@HMOS catalyst [(E-G), adapted with permission from Kuwahara et al. (*119*)]. (H-J) Electrocatalytic conversion of $CO_2$ into formate by $Pd_1Cu$ catalyst: (H) schematic illustration of $CO_2$ conversion into HCOOH over a $Pb_1Cu$ catalyst, (I) FEs of all $CO_2RR$ products at different current densities and the corresponding j-V curve of $Pb_1Cu$ catalyst, (J) theoretical calculation of adsorption free energies for HCOO* and COOH* and free-energy diagrams for the competition between $CO_2RR$ producing formate or CO and the HER reaction [(H-J), adapted with permission from Zheng et al. (*121*)].

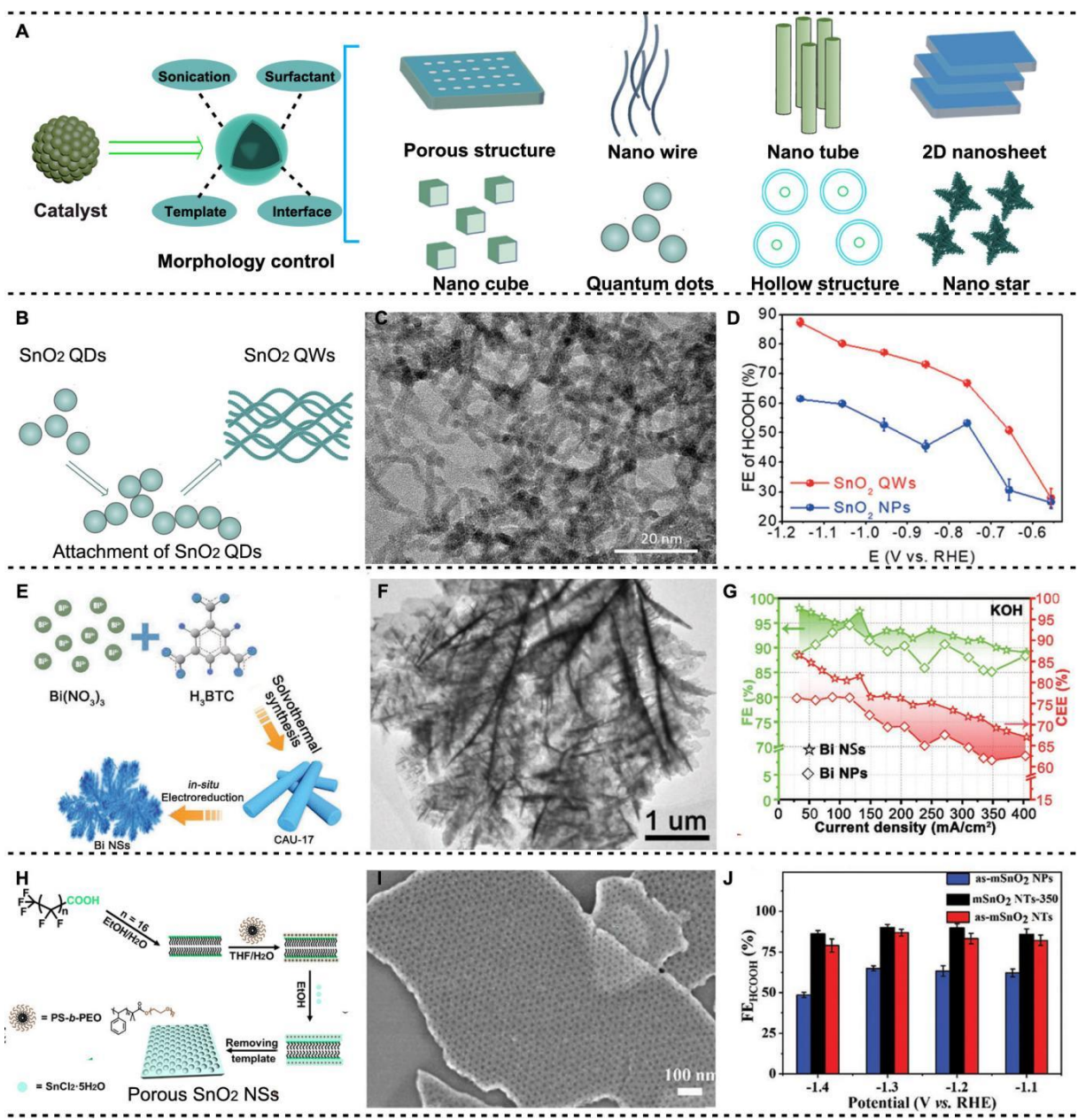

**Figure 5. Conversion of CO$_2$ into carboxylic acids by morphology control.** (A) Schematic diagram of tuning the catalytic activity of catalysts using morphology control. (B-D) Electrocatalytic conversion of CO$_2$ into formate with SnO$_2$ QWs: (B) Illustration of the structure of ultrathin SnO$_2$ QWs, (C) TEM image of SnO$_2$ QWs, (D) FE of HCOOH over the ultrathin SnO$_2$ QWs and SnO$_2$ nanoparticles [(B-D), adapted with permission from Liu et al. (*134*)]. (E-G) Electrocatalytic conversion of CO$_2$ into formate with Bi nanosheets (NSs): (E) Schematic diagram of the synthesis of Bi NSs and illustration of the structure of Bi NSs, (F)

TEM image of Bi NSs, (G) FEs and cathodic energetic efficiency (CEE) of formic acid over two electrocatalysts in 1 m of KOH [(E-G), adapted with permission from Yang et al. (*135*)]. H-J) Electrocatalytic conversion of $CO_2$ into formate with porous $SnO_2$ NSs: (H) Schematic diagram of the synthesis of porous $SnO_2$ NSs and illustration of the structure of porous $SnO_2$ NSs, (I) SEM image of porous $SnO_2$ NSs, (J) Faradaic efficiencies of HCOOH over different catalysts [(H-J), adapted with permission from Wei et al. (*136*)].

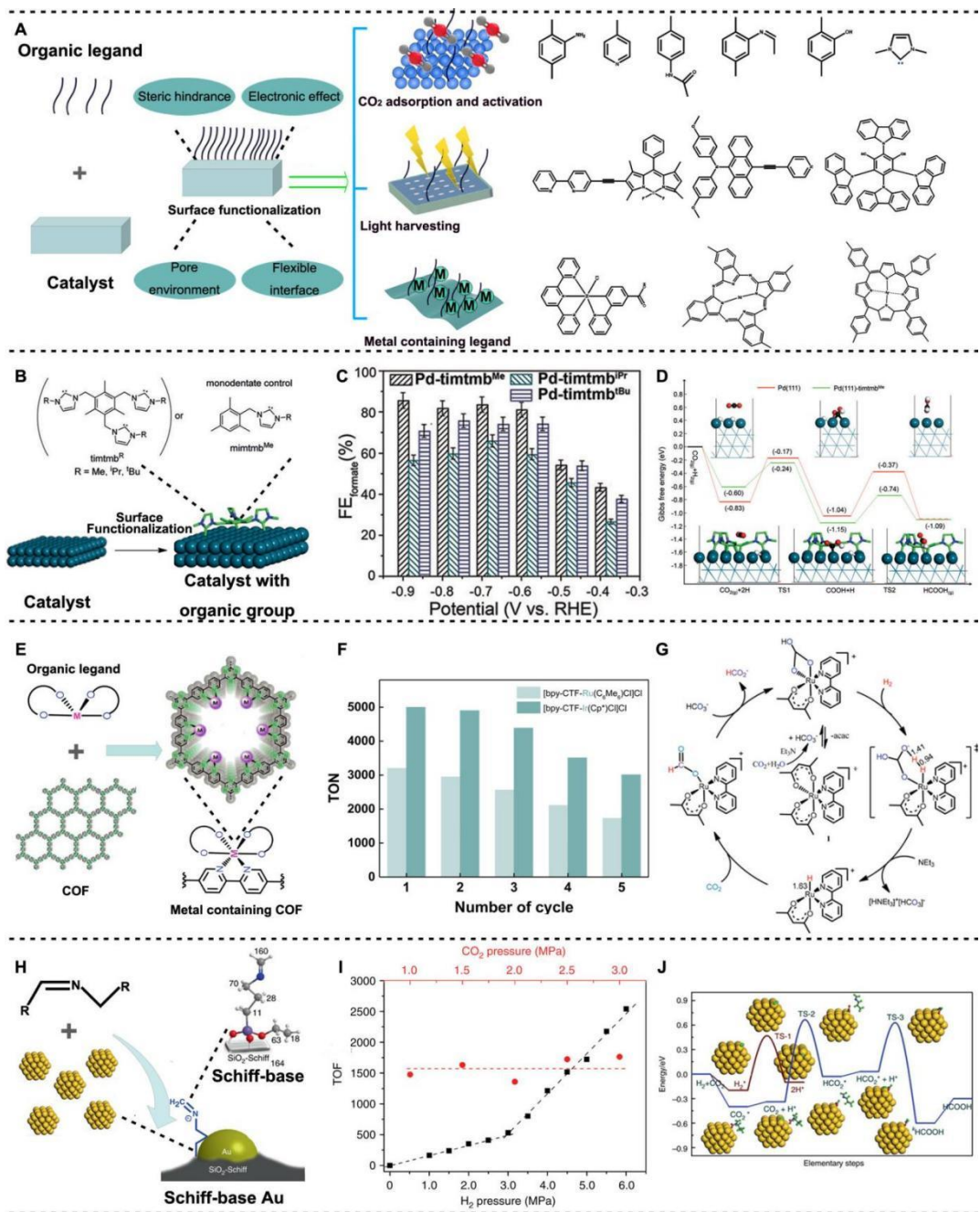

**Figure 6. Conversion of CO$_2$ into carboxylic acids by surface functionalization.** (A) Schematic diagram of tuning catalytic activity via surface functionalization. (B-D) Electrocatalytic conversion of CO$_2$ into formate with tripodal N-heterocyclic carbene functionalized palladium: (B) Synthetic scheme of tripodal N-heterocyclic carbene functionalized palladium, (C) FEs of formate generation using Pd-timtmb$^R$ electrodes, (D) Free-energy diagrams of CO$_2$ reduction to HCOOH on Pd(111) and Pd(111)-timtmb$^{Me}$ [(B-D),

adapted with permission from Cao et al. (*150*)]. (E-G) Hydrogenation of $CO_2$ with [bpy-CTF-Ru(acac)$_2$]Cl: (E) Schematic diagram of [bpy-CTF-Ru(Ic)$_2$]Cl synthesis, (F) reusability tests of [bpy-CTF-Ru(acac)$_2$]Cl for $CO_2$ hydrogenation, (G) proposed mechanism for hydrogenating $CO_2$ into formate [(E-G), adapted with permission from Gunasekar et al. (*151*)]. (H-J) Hydrogenation of $CO_2$ with a Schiff-base-modified gold nanocatalyst: (H) Schematic diagram of the Au/SiO$_2$-Schiff structure (insert: quantum mechanics calculation of NMR shifts), I) initial reaction rates in $H_2$ and $CO_2$ pressure-dependent conditions over Au/SiO$_2$-Schiff, (J) Free-energy diagram of $CO_2$ hydrogenation over a Au/SiO$_2$-Schiff catalyst [(H-J), adapted with permission from Liu et al. (*153*)].

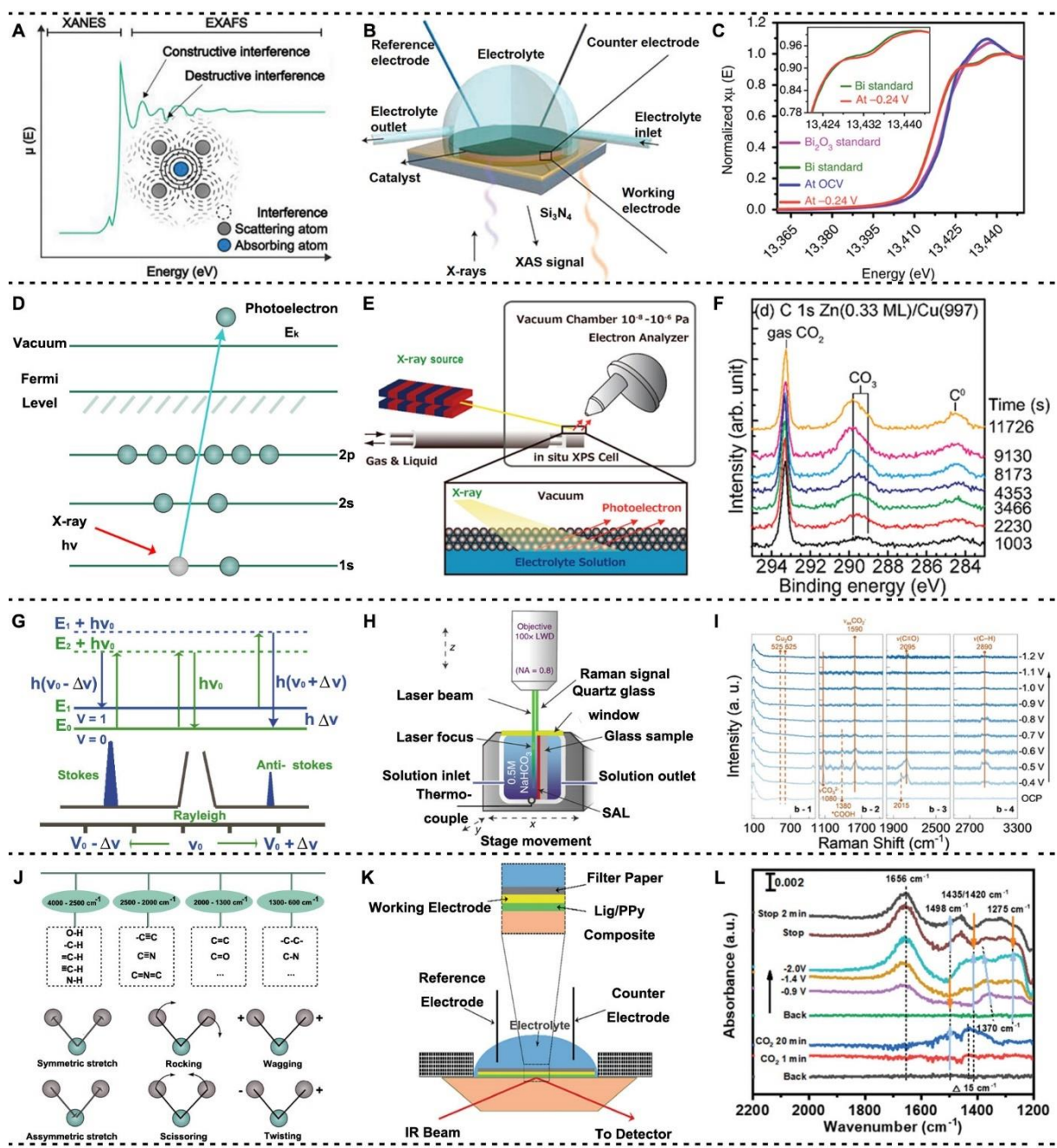

**Figure 7. In situ characterization techniques.** (A) Schematics of the x-ray absorption spectra (XAS), adapted with permission from Long et al. (*157*). (B) Schematic diagram of operando XAS cell, adapted with permission from Zheng et al. (*185*). (C) Operando Bi L-edge XANES spectra of $Bi_2O_3$ nanotubes and NTD-Bi at -0.24 V compared to Bi or $Bi_2O_3$ standards (inset plot: partially enlarged spectra), adapted with permission from Gong et al. (*158*)) (D) Schematic diagram of the XPS principle. (E) Schematic diagram of operando XPS cell, adapted

with permission from Li et al. (*186*)). (F) In situ x-ray photoelectron spectrum over Zn (0.33 ML)/Cu (997) in the presence of 0.8 mbar $CO_2$ + 0.4 mbar $H_2$, adapted with permission from Koitaya et al. (*159*). (G) Schematic diagram of the Raman spectrum principle. (H) Schematic diagram of the operando Raman cell, adapted with permission from Geisler et al. (*187*). (I) in situ Raman spectra with a static scanning subsection of the S-doped catalyst, adapted with permission from Pan et al. (*160*). (J) Schematic diagram of Fourier-transform infrared (FTIR) spectroscopy principle. (K) Schematic diagram of an in situ FTIR spectroscopy cell, adapted with permission from Ajjan et al. (*188*). (L) In situ attenuated total reflection surface-enhanced infrared absorption spectroscopy on Sn-OH-5.9, adapted with permission from Deng et al. (*161*).

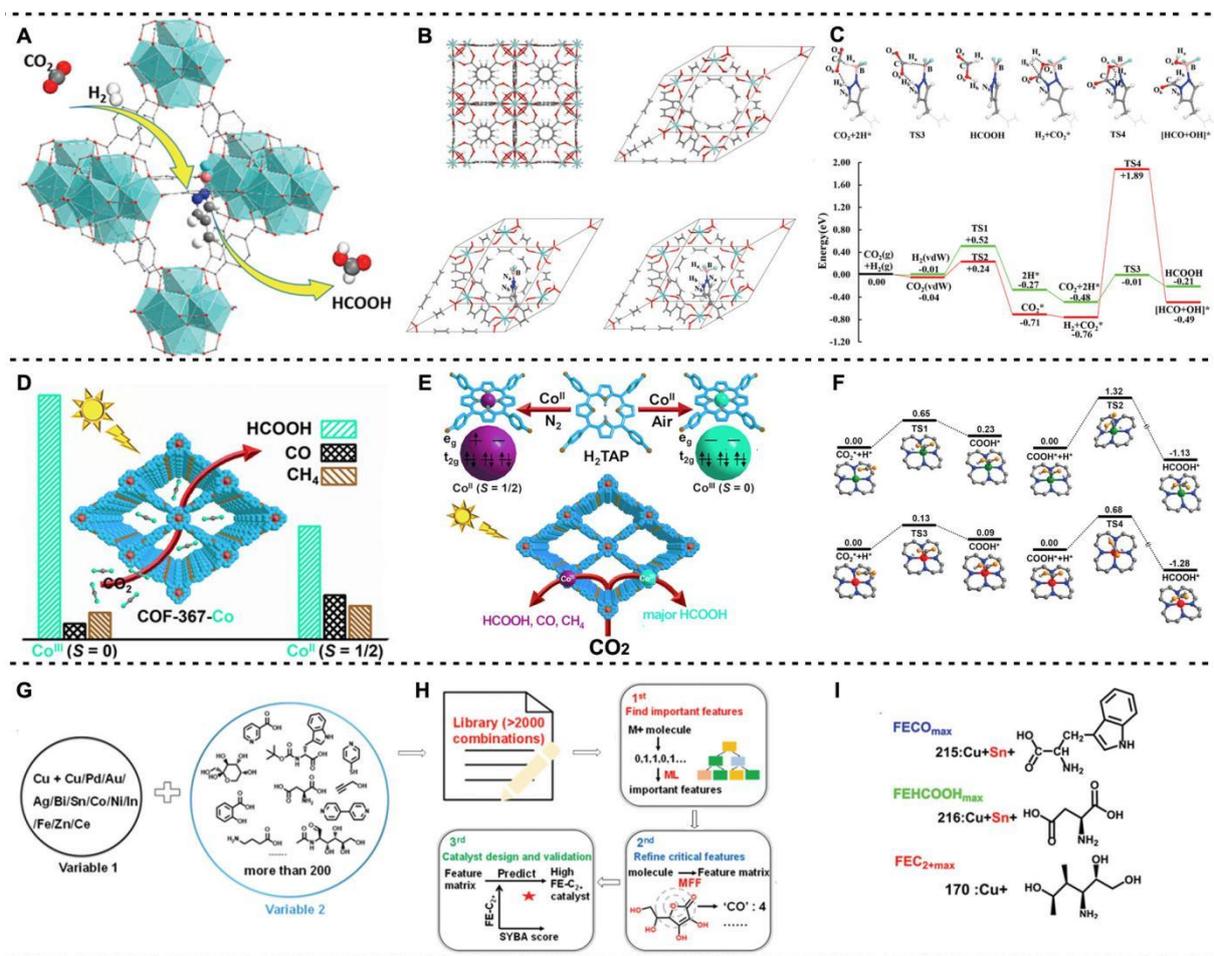

**Figure 8. Theoretical calculations.** (A-C) $CO_2$) hydrogenation over UiO-66-P-BF$_2$: (A) Schematic diagram of $CO_2$ hydrogenation over UiO-66-P-BF$_2$, (B) schematic diagram of the structure of UiO-66-P-BF$_2$, (C) relative potential energy surfaces for two $CO_2$ hydrogenation pathways in UiO-66-P-BF$_2$ and the involved configuration in the process of $CO_2$ hydrogenation [(A-C), adapted with permission from Ye et al. (*167*)]. (D-F) Photocatalytic conversion of $CO_2$ into formic acid over COF-367-Co: (D) Schematic diagram of photocatalytic conversion of $CO_2$ into formic acid over COF-367-Co, (E) schematic diagram of COF-367-Co featuring different spin states of Co ions toward photocatalytic $CO_2$ reduction, (F) calculated potential energy profile of $CO_2$ reduction reaction to HCOOH catalyzed by COF-367-Co$^{II}$ and COF-367-Co$^{III}$ [(D-F), adapted with permission from Gong et al. (*168*)]. (G-I) Machine-learning-

guided discovery and optimization of additives in preparing Cu catalysts for $CO_2$ reduction: (G) Schematic diagram of combinations of metal salts and additives, (H) schematic diagram of the process of high-performance catalysts under the guide of machine learning, (I) Additives screened out by machine learning [(G-I), adapted with permission from Guo et al. (*171*)].

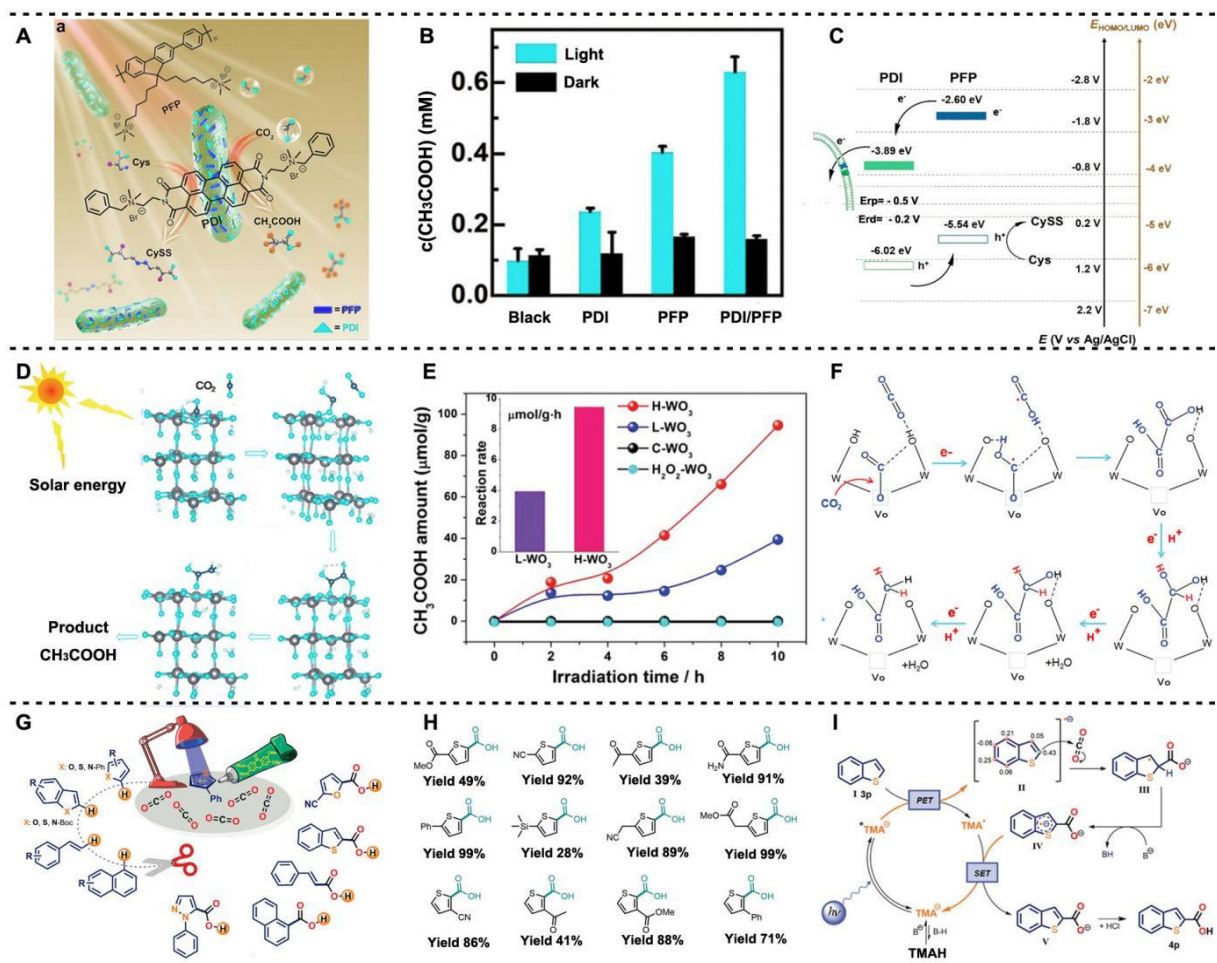

**Figure 9. Frontier progress of photoreduction of $CO_2$ into carboxylic acid.** (A-C) Photocatalytic conversion of $CO_2$ into acetic acid: (A) Schematic diagram of the photosynthetic production of acetic acid by conjugated molecules/M. thermoacetica, (B) produced acetic acid amount of PDI/PFP/M. thermoacetica in an alternating light-dark cycle of 12 h each, (C) energy-level diagram of PDI/PFP and the representation of the electron transfer mechanism to the bacterial membrane [(A-C), adapted with permission from Gai et al. (*173*)]. (D-E) Photocatalytic conversion of $CO_2$ into acetic acid over $WO_3 \cdot 0.33H_2O$: (D) Schematic diagram of photocatalytic conversion of $CO_2$ into acetic acid over $WO_3 \cdot 0.33H_2O$, (E) catalytic results with irradiation time over different catalysts, (F) photocatalytic conversion of $CO_2$ with $WO_3 \cdot 0.33H_2O$ [(D-F), adapted with permission from Sun et al. (*174*)]. (G-I) Redox-neutral

photocatalytic C-H carboxylation with $CO_2$: (G) Schematic diagram of photocatalytic carboxylation, (H) catalytic results of various substrates using photocatalytic conversion, (I) proposed mechanism for the C-H carboxylation of (hetero)arenes [(D-F), adapted with permission from Schmalzbauer et al. (*175*)].

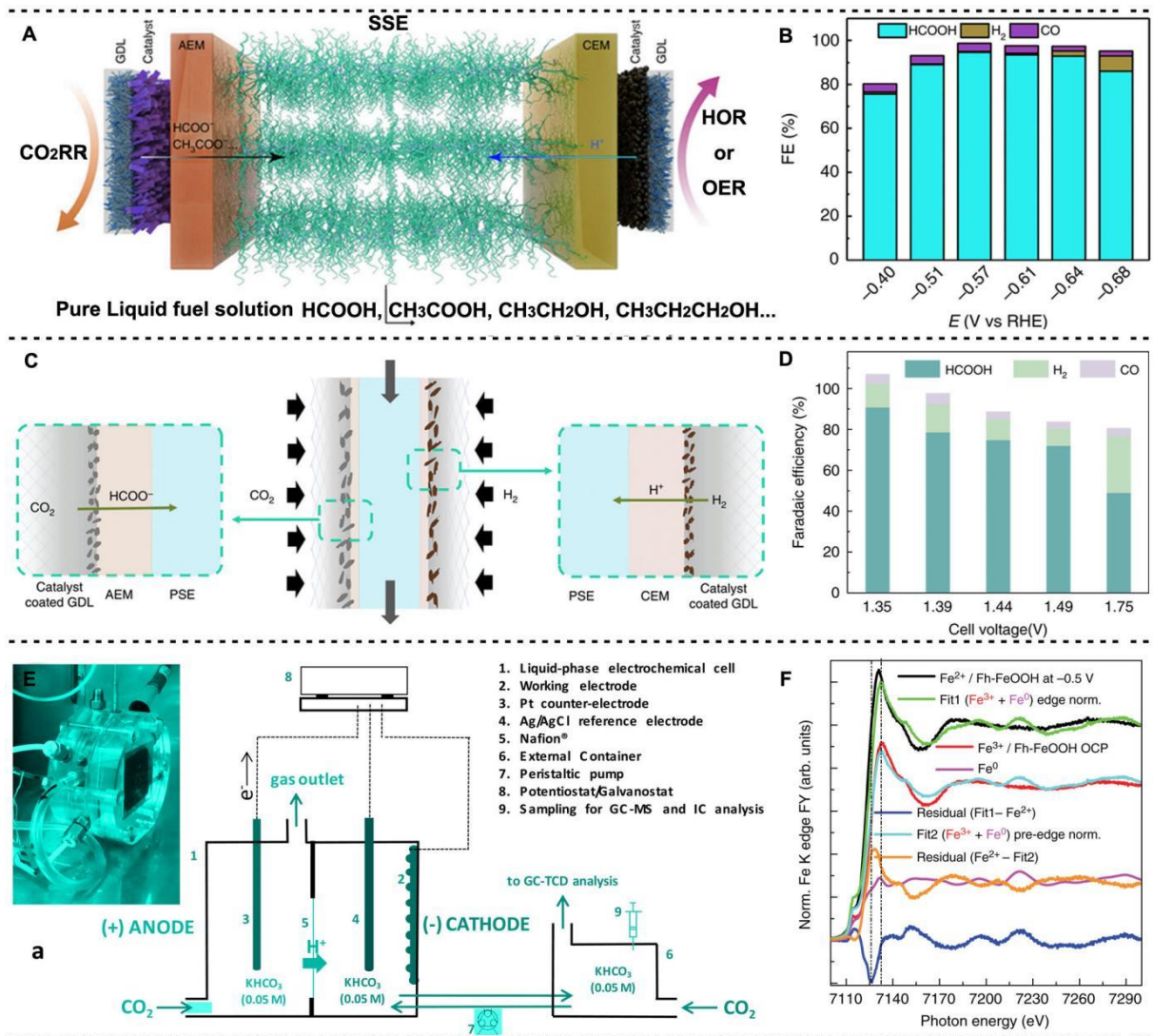

**Figure 10. Frontier progress of electroreduction of $CO_2$ into carboxylic acid.** (A-B) Electrocatalytic conversion of carbon dioxide ($CO_2$) into formic acid with solid electrolytes: (A) Schematic illustration of the $CO_2$ reduction cell with solid electrolytes, (B) corresponding FEs of the resultant reduction products [(A-B), adapted with permission from Xia et al. (*176*)]. (C-D) Electrochemical $CO_2$ reduction to a high concentration of pure formic acid solutions in an all-solid-state reactor: (C) Schematic illustration of the all-solid-state electrochemical $CO_2$RR to the formic acid reactor (AEM: anion exchange membrane, CEM: cation exchange membrane, GDL: gas diffusion layer, PSE: porous solid electrolyte), (D) corresponding FEs of formic acid vapor under different cell voltages [(C-D), adapted with permission from Fan et al. (*177*)]. (E-F) Operando spectroscopy study of $CO_2$ electroreduction by iron species on nitrogen-doped carbon: (E) Schematic drawing of the experimental cell with details of electrodes, charge pathways, and electrolytes, (F) in situ normalized Fe k-edge spectra of Fe/N-C [(E-F), adapted with permission from Genovese et al. (*180*)].

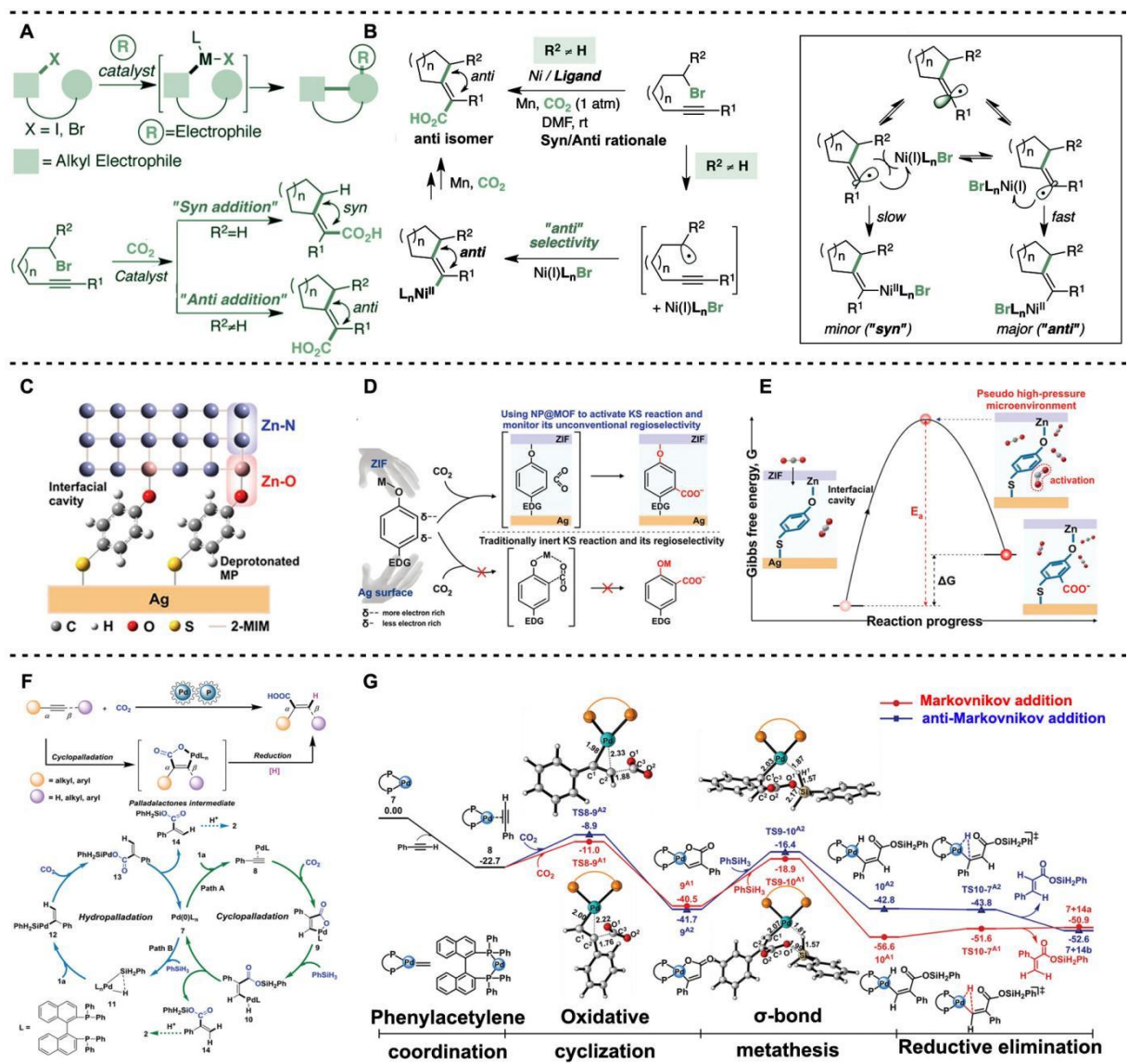

**Figure 11. Frontier progress of thermo-catalytic conversion $CO_2$ into carboxylic acid.** (A-B) Ni-catalyzed divergent cyclization/carboxylation of unactivated primary and secondary alkyl halides with carbon dioxide ($CO_2$): (A) Schematic diagram of bond formation via electrophile couplings and cyclization/functionalization of alkyl halides, (B) mechanism of catalytic cyclization/carboxylation of unactivated secondary alkyl halides [(A-B), adapted with permission from Wang et al. (*182*)]. (C-E) Production of carboxylic acids using Kolbe–Schmitt (KS) reactions: (C) Proposed molecular configuration at the Ag-MP@ZIF interface and Zn 2p XPS spectra of Ag-MP@ZIF and Ag-MBT@ZIF, (D) schematic depicting the importance of the NP@MOF interface to activate an inert KS reaction and in situ monitor its unconventional regioselectivity, E) proposed mechanism for how NP@MOF activates a chemical reaction that is otherwise inert at ambient operating conditions [(C-E), adapted with permission from Lee et al. (*183*)]. (F-G) Pd-catalyzed highly regioselective hydrocarboxylation of alkynes with $CO_2$: (F) Schematic diagram and mechanism of hydrocarboxylation of alkynes with $CO_2$, (G) calculated relative Gibbs free energies of catalytic species through the cyclopalladation pathway [(F-G), adapted with permission from Xiong et al. (*184*)].